\documentclass[12pt]{article}
\usepackage{amsfonts,amssymb,amsmath,epsfig,graphicx}
\usepackage{theorem,calc}
\newtheorem{definition}{Definition}[section]

\newtheorem{conjecture}[definition]{Conjecture}
\newtheorem{proposition}[definition]{Proposition}

\newtheorem{theorem}[definition]{Theorem}

\newtheorem{rmk}{Remark}[section]

\numberwithin{equation}{section}

\newcommand{\nonu}{\nonumber \\}

\newcommand{\hs}[1]{\hspace{#1 mm}}

\newcommand{\vph}{\varphi}

\def\cA{{\cal A}}            
\def\cD{{\cal D}}      \def\cE{{\cal E}}      
      \def\cH{{\cal H}}      
      \def\cK{{\cal K}}      \def\cL{{\cal L}}
\def\cM{{\cal M}}      \def\cN{{\cal N}}      
            \def\cR{{\cal R}}
\def\cS{{\cal S}}            \def\cU{{\cal U}}

\def\fc{{\mathfrak c}}

\def\fm{{\mathfrak m}}
\def\fn{{\mathfrak n}}
\def\fp{{\mathfrak p}}
\def\fr{{\mathfrak r}}

\newcommand{\CC}{{\mathbb C}}

\newcommand{\II}{{\mathbb I}}

\newcommand{\RR}{\mbox{${\mathbb R}$}}

\newcommand{\ZZ}{{\mathbb Z}}

\newcommand{\prt}{\partial}

\newcommand{\wh}[1]{\widehat{#1}}
\newcommand{\wt}[1]{\widetilde{#1}}
\newcommand{\mb}[1]{\hs{4}\mbox{#1}\hs{4}}
\newcommand{\qmbox}[1]{{\qquad\mbox{#1}\quad}}
\newcommand{\half}{\frac{1}{2}}

\newcommand{\prf}{\underline{Proof:}\ }
\newcommand{\finprf}{\null \hfill {\rule{5pt}{5pt}}\\[2.1ex]\indent}

\newcommand{\atopn}[2]{\genfrac{}{}{0pt}{}{#1}{#2}}

\newcommand{\bigchk}[1]{{\stackrel{\scriptscriptstyle\vee}{#1}}}
\newcommand{\de}{\rho}

\newcounter{heure}
\newcounter{minute}

\begin{document}
\renewcommand{\thefootnote}{\arabic{footnote}}
\setcounter{footnote}{0}
\newpage
\setcounter{page}{0}

\pagestyle{empty}

\null
\vfill
\begin{center}

{\Large \textsf{Scattering matrix for a general 
$gl(2)$ spin chain}}

\vspace{10mm}

{\large S. Belliard$^a$, N. Cramp{\'e}$^{b}$ and
{\'E}. Ragoucy$^a$}
\footnote{samuel.belliard@lapp.in2p3.fr, nicolas.crampe@lpta.univ-montp2.fr, 
eric.ragoucy@lapp.in2p3.fr}

\vspace{10mm}
\emph{$^a$ Laboratoire d'Annecy-le-Vieux de Physique Th{\'e}orique}

\emph{LAPTH, CNRS et Universit{\'e} de Savoie, UMR 5108}

\emph{B.P. 110, F-74941 Annecy-le-Vieux Cedex, France}
\vspace{5mm}\\

\emph{$^b$ Laboratoire de Physique Th{\'e}orique \& Astroparticules}

\emph{LPTA, CNRS, UMR 5207, Universit{\'e} Montpellier II}

\emph{F-34095, Montpellier Cedex 5, France}
\vspace{5mm}\\
\vfill

\end{center}

\vfill
\vfill

\begin{abstract}
We study the general $L_0$-regular $gl(2)$ spin chain, i.e. a 
chain where the sites 
$\{i,i+L_0,i+2L_0,\dots\}$ carry the same arbitrary representation 
(spin) of 
$gl(2)$. The basic example of such chain is obtained 
for $L_0=2$, where we recover the alternating spin chain. \\
Firstly, we review different 
known results about their integrability 
and their spectrum.
Secondly, we give an interpretation in terms of particles and 
conjecture the scattering matrix between them.
\end{abstract}

\vfill
\centerline{
MSC: 81R50, 17B37 ---
PACS: 02.20.Uw, 03.65.Fd, 75.10.Pq}
\vfill

\rightline{\texttt{arXiv:0909.1520 [math-ph]}\qquad}
\rightline{LAPTH-1351/09\qquad}
\rightline{PTA/09-059\qquad}
\rightline{September 2009\qquad}

\baselineskip=16pt

\newpage

\newpage
\pagestyle{plain}

\section{Introduction}

The one-dimensional Heisenberg spin chain \cite{heisen} is among the few 
many-body quantum systems for which
one can compute exactly some physical quantities (spectrum, 
correlation functions, ...). This model 
was solved for the first time in the seminal paper of H. Bethe 
\cite{bethe} where he succeeded to map 
the study of the spectrum to the resolution of transcendental 
equations, now called Bethe equations.    
Subsequently, the Quantum Inverse Scattering Method (QISM) has been 
introduced (see \cite{kusk,tafa} and for reviews \cite{KoIzBo,fafa}) 
based on solutions of 
Yang-Baxter equation \cite{yang,baxter}. This approach is very 
fruitful and, for example, provides a whole 
class of integrable spin chains associated to generic algebras (Yangians 
or quantum groups based on classical Lie algebras or superalgebras). 
Focusing on spin chains based on $gl(2)$ only, one can use the QISM 
approach to define and study e.g.: 
higher spin chains \cite{ZAFA,Bab}, 
spin chain with impurities \cite{anjo,YWang} 
or alternating spin chain \cite{dewo,abad2,ana}. In these types of 
chains the spins are no more 1/2, but can be arbitrary, a situation 
that is nowadays relevant for condensed matter experiments, where 
quasi-one-dimensional spin chains with different spins are studied, 
see e.g. \cite{condmat}. 

It is thus natural to wonder whether a spin chain containing arbitrary 
spins can be studied through QISM approach. A first and immediate 
problem in such studies comes from the thermodynamical limit (when 
$L$, the number of sites, tends to infinity). Such limit is 
needed for a comparison with physical model, and it is rather obvious that 
one needs some regularity in the spin content of the chain so as to be 
able to compute relevant quantities while taking the limit $L\to\infty$. 

The $L_0$-regular 
spin chains (i.e. spin chains with a repeated motif containing $L_0$ 
spins in arbitrary representations) have been introduced 
\cite{ow,thermy} to remedy 
this objection while keeping enough freedom to encompass most of the
 known cases. For instance, homogeneous spin chains (for any spin $s$) 
correspond to $L_{0}=1$, and 
one recovers the alternating spin chain for $L_0=2$. They also allow 
to define an integrable model \cite{thermy} for spin chains  with 
periodic array of impurities (while the original ones \cite{fuka} 
were not integrable).  $L_0$-regular spin chains 
have been also studied using QISM approach in 
\cite{ow,byebye,thermy,therm2}. 

In the antiferromagnetic regime, these models may be seen as lattice 
versions of a corresponding integrable 
relativistic quantum field theories. This link is 
very useful since it allows one to 
compare both models for which numerous exact results are known. 
{From} spin chains side, by solving the Bethe equations
in the thermodynamical limit, it may be possible to compute exactly the
scattering matrix between the excitations. It allows one to obtain 
indication on the underlying field theory and, 
then, on the long-distance physics.
This program has been followed for different types of spin chain: 
Heisenberg model \cite{faddeev}, 
higher spin chains \cite{Tak,NRe} and alternating spin $(1/2,1)$ 
chain \cite{dev,mene}. In this paper, 
we tackle the problem to compute the scattering matrix for the 
general $L_0$-regular spin chains based on $gl({2})$. 
After recalling some results about their spectrum, we give an 
interpretation of 
the excitations in terms of particles and conjecture an explicit form 
for the scattering matrix.

The outline of this paper is as follows. In section \ref{sec:not}, we 
give the notations used throughout 
the paper. Then, we recall, in section \ref{sec:Hint}, well-known 
results about the integrability of the 
general spin chains using the transfer matrices 
constructed from rational solutions of the Yang-Baxter equation. 
We also link these transfer matrices 
with the shift operator as well as the 
Hamiltonian of the models. Section 
\ref{sec:bet} is devoted to the computation of the Bethe equations 
and their study in the string hypothesis. 
The two following sections give some results on the spectrum: the 
energy of the antiferromagnetic vacuum is 
given in section \ref{sec:vac} and the dispersion law for the first 
excited states is established in
section \ref{sec:exct}. Then, we propose, in section \ref{sec:Smat}, 
a conjecture for the scattering matrix between these excited states.
Finally, in section \ref{sec:conclu}, we conclude on open problems.

\section{Notation \label{sec:not}}

%
\paragraph{$\mathbf{gl(2)}$ Lie algebra}

We introduced the spin $s$ representation of $gl(2)$ given explicitly 
by
\begin{eqnarray}
\pi_{s}(e_3) = \sum_{n=1}^{2s+1}\big(s+1-n\big)\,E^{(s)}_{nn}
\mb{;}
\pi_{s}(e_{+}) = \sum_{n=1}^{2s}\sqrt{n(2s+1-n)}
\,E^{(s)}_{n,n+1}
\nonu
\pi_{s}(e_{-}) = \sum_{n=1}^{2s} \sqrt{n(2s+1-n)}
\,E^{(s)}_{n+1,n}
\mb{;}
\pi_{s}(e_0) = \sum_{n=1}^{2s+1} E^{(s)}_{nn} = \II_{2s+1}
\end{eqnarray}
where $E_{nm}^{(s)}$ is a $(2s+1)\times (2s+1)$ matrix with $1$ in the
entry $(n,m)$ and $0$ otherwise.
The spin $s$ representation of $su(2)$ embedded in $gl(2)$ is 
generated by 
$\{ \pi_{s}(e_3),\pi_{s}(e_+),\pi_{s}(e_-)\}$.

\paragraph{Set of representations}

We study a periodic $gl(2)$ spin chain of $L$ sites
with the spin $s_i$ representation on the site $i$.
To be able to take the thermodynamical limit, we restrict ourselves 
to the case
of $L_0$-regular spin chain (i.e. $s_i=s_{i+L_0}$). In this case, the 
length $L$ of 
the chain must be chosen such
that $L/L_0$ be an integer.
We introduce the ordered set
$\cS=\{\bar s_1,\bar s_2,\dots\bar s_\cL\,|\ \bar s_j<\bar s_{j+1}\}$
of the different values of the spins $s_i$ ($1\leq i \leq L_0$) 
present in the spin 
chain.
We denote by $L_{\bar s_j}$ the number of times $\bar s_j$ appears in 
the sequence
$s_1,s_2,\dots,s_{L_0}$ which allows us to define the density of the 
spin
$\bar s_j$ in the chain by
\begin{equation}
\de_{\bar s_j}=L_{\bar s_j}/L_0\;.
\end{equation}
We get $\de_{\bar s_1}+\dots+\de_{\bar s_\cL}=1$.
For convenience, we use the conventions $\bar s_0=0$ and $\bar 
s_{\cL+1}=\infty$.

We will also need to consider the spins which are \underline{not}
present in the chain, so that we introduce the following sets, for 
$i=0,1,\dots,\cL$,
\begin{equation}
\label{eq:Rs}
\cR_{\bar s_i}=\{\bar s_i+\half\,,\,\bar s_i+1\,,\,\dots,\bar 
s_{i+1}-\half\}
\, =\ ]\bar s_i\,,\,\bar s_{i+1}[\ \cap\,\half\,\ZZ\;.
\end{equation}
We define  $\displaystyle \cR=\bigcup_{i=0}^\cL\cR_{\bar 
s_i}=\half\,\ZZ_{>0}
\setminus \cS$ which is
the set of all the representations not used to construct the spin 
chain.

Finally, to make lighter the formulas, we will use sometimes
$\de_j$ (resp. $\cR_j$) instead of $\de_{\bar s_j}$ (resp. $\cR_{\bar 
s_j}$).
For instance, $\cR_{0}=\{\half\,,\,1\,,\,\dots,\bar s_1-\half\}$ and 
$\cR_{\cL}=\{\bar s_\cL+\half\,,\,\bar s_\cL+1\,,\,\dots,\infty\}$.

\paragraph{Elementary functions}

In the whole paper, essentially two functions as well as their 
logarithm, their
derivative and their Fourier transform are necessary to construct all 
the other ones.
We use the following definition for the Fourier transform
\begin{equation}
\hat f(p)=\frac{1}{2\pi}\int_{-\infty}^\infty e^{ip\lambda} 
f(\lambda) d\lambda\;.
\end{equation}
We encompass in  figure \ref{fig:elemfct} their explicit form and 
their relations,
for $\hbar>0$ and $0<r<\frac{\pi}{\hbar}$.
\begin{figure}[htb]
\begin{equation*}
\begin{array}{c c c c}
\displaystyle
G_r^{(\hbar)}(\lambda)=\frac{\displaystyle
\sinh\left(\hbar\left(-\lambda-\frac{ir}{2}\right)\right)}
{\sinh\left(\displaystyle\hbar\left(\lambda-\frac{ir}{2}\right)\right)}
&
\hspace{5mm}\xrightarrow{\displaystyle\ \ \lim_{\hbar\rightarrow 0}\ 
\ }
&
\displaystyle
-e_r(-\lambda)=\frac{\displaystyle-\lambda-\frac{ir}{2}}
{\displaystyle\lambda-\frac{ir}{2}} & \\[2.1ex]
\downarrow && \downarrow & i\ln(.)\\[2.1ex]
\displaystyle \Gamma_r^{(\hbar)}(\lambda)=
2\arctan\Big(\ \frac{\tanh\left(\hbar\lambda\right)}
{\displaystyle\tan\left(\frac{\hbar r}{2}\right)}\ \Big)
&&
\displaystyle\varphi_r(\lambda)=2\arctan\left(\frac{2\lambda}{r}\right)
&\\[2.1ex]
\downarrow && \downarrow & \text{Derivative w.r.t. $\lambda$}\\[2.1ex]
\displaystyle
\gamma_r^{(\hbar)}(\lambda)=\frac{2\hbar\sin(\hbar r)}{\cosh(2\hbar 
\lambda)-
\cos(\hbar r)}
&&\displaystyle\frac{4r}{4\lambda^2+r^2}&
\\[4.2ex]
\downarrow && \downarrow & \text{Fourier transform}\\[2.1ex]
\displaystyle \wh\gamma_r^{(\hbar)}(p)=
\frac{ \sinh
\left[\displaystyle\frac{p}{2}\left(\frac{\pi}{\hbar}-r\right)\right]}
{\displaystyle\sinh\left(\frac{p\pi}{2\hbar}\right)}
&&\displaystyle \exp\left(\displaystyle-\frac{r|p|}{2}\right) &
\end{array}
\end{equation*}
\caption{Relations between the elementary functions used in the
paper\label{fig:elemfct}}
\end{figure}

We extend these definitions  to degenerate cases in the
following way. For $r=0$:
\begin{equation}
\label{eq:de1}
G_0^{(\hbar)}(\lambda)=-1\qmbox{;}\Gamma_0^{(\hbar)}(\lambda)=0\qmbox{;}
\gamma_0^{(\hbar)}(\lambda)=2\pi\delta(\lambda)\qmbox{and}
\wh\gamma_0^{(\hbar)}(p)=1\;,
\end{equation}
and, when $r=\frac{\pi}{\hbar}$,
\begin{equation}
\label{eq:de2}
G_{\pi/\hbar}^{(\hbar)}(\lambda)=1\qmbox{;}
\Gamma_{\pi/\hbar}^{(\hbar)}(\lambda)=0\qmbox{;}
\gamma_{\pi/\hbar}^{(\hbar)}(\lambda)=0\qmbox{and}
\wh\gamma_{\pi/\hbar}^{(\hbar)}(p)=0\;.
\end{equation}

We need also the more involved following functions defined, for  
$0<r<\frac{\pi}{\hbar}$, by
\begin{equation}
\wh\kappa_{r}^{(\hbar)}(p)
=\frac{\wh\gamma_{r}^{(\hbar)}(p)}{2\cosh(\frac{p}{2})}
\;,
\end{equation}
and
\begin{equation}
\cK_{r}^{(\hbar)}(\lambda)=\exp\int_{-\infty}^{\infty}dp\ 
\frac{e^{-ip\lambda}}{p}\ \wh\kappa_{r}^{(\hbar)}(p)
=\exp-i \int_{0}^{\infty}dp \frac{\sin(p\lambda)}{p}\ 
\frac{\sinh(\frac{p}{2}(\frac{\pi}{\hbar}-r))}
{\cosh(\frac{p}{2})\sinh(\frac{p\pi}{2\hbar})}\;.
\end{equation}
We extend also the previous definitions to the cases $r=0$ and 
$r=\pi/\hbar$ 
using the conventions:
\begin{equation}
\label{eq:exK}
\cK_{0}^{(\hbar)}(\lambda)
=-i\coth\left(\frac{\pi}{2}\left(\lambda-\frac{i}{2}
\right)\right)
\qmbox{and}
\cK_{\pi/\hbar}^{(\hbar)}(\lambda)=1\;.
\end{equation}
The limit $\hbar\rightarrow 0$ of $\cK_{r}^{(\hbar)}(\lambda)$ can be 
computed 
and we get
\begin{equation}
\cK_{r}^{(0)}(\lambda)=\frac{\Gamma(-\frac{i\lambda}{2}+\frac{r+3}{4})
\Gamma(\frac{i\lambda}{2}+\frac{r+1}{4})}
{\Gamma(\frac{i\lambda}{2}+\frac{r+3}{4})
\Gamma(-\frac{i\lambda}{2}+\frac{r+1}{4})}\;.
\end{equation}
The limits at $\pm \infty$ will be also used in the following, for 
$0\leq r \leq \frac{\pi}{\hbar}$,
\begin{equation}
\label{eq:limits}
\lim_{\lambda\rightarrow\pm\infty} G_{r}^{(\hbar)}(\lambda)=
\exp(\mp i(\pi-\hbar r))
\quad\text{and}\qquad
\lim_{\lambda\rightarrow\pm\infty} \cK_{r}^{(\hbar)}(\lambda)=
\exp(\mp\frac{i}{2}(\pi-\hbar r))\;.
\end{equation}

\section{Integrable Hamiltonians\label{sec:Hint}}
\subsection{Monodromy and transfer matrices}

We will need monodromy matrices of different types, depending on 
the
auxiliary space representation. Indeed, for $i=1,\ldots,\cL$, we
define the monodromy matrix with auxiliary space
in the spin $\bar s_i$ representation as
\begin{equation}
T^{(\bar s_i)}_0(u)=\prod_{0\leq p<L/L_0}^{\longrightarrow}
R^{(\bar s_i,s_1)}_{0,1+pL_0}(u)\dots
R^{(\bar s_i,s_{L_0})}_{0,(p+1)L_0}(u)
\end{equation}
where the product is ordered
$\displaystyle \prod_{0\leq i< 
L/L_0}^{\longrightarrow}X_{1+i}=X_1X_2\dots X_{L/L_0}$.
 $R^{(s_i,s_j)}_{0,j}(u)$ may be obtained by fusion 
\cite{Kul}.
We do not recall here their construction and 
give only their explicit form
\begin{equation}
R^{(s,s')}(u)=
\sum_{k=|s-s'|}^{s+s'}f^{(s,s')}_k(u) \mathcal{P}^{(s,s')}_k
\end{equation}
with $\displaystyle
f^{(s,s')}_k(u)=
\prod_{\ell=k+1}^{s+s'}\left(\frac{u-i\ell}{u+i\ell}\right)
$. As usual, one have introduced the following projectors
\begin{equation}
\mathcal{P}^{(s,s')}_k=\prod_{\genfrac{}{}{0pt}{}{j=|s-s'|}{j\neq 
k}}^{s+s'}
\frac{(\pi_s\otimes\pi_{s'})(e_3\otimes e_3+\frac{1}{2}
(e_+\otimes e_- + e_-\otimes e_+))-x_j}{x_k-x_j}
\end{equation}
with $x_k=\frac{1}{2}[k(k+1)-s(s+1)-s'(s'+1)]$.
In particular, for $s=s'=\half$, we get the usual Yang's R-matrix 
\cite{yang}
\begin{equation}
\label{eq:RR}
R_{12}(u)=R^{(\half,\half)}_{12}(u)=\frac{1}{u+i}(u+i P_{12})\;.
\end{equation}

The normalization has been chosen such 
that it leads to regular and unitary
matrices:
 \begin{equation}
 \label{eq:prfused}
 R^{(s,s)}_{0,i}(0)=P_{0,i}^{(s)}
  \qmbox{and}
 R^{(s_i,s_j)}_{0,j}(u)R^{(s_i,s_j)}_{0,j}(-u)=1
 \end{equation}
where $P_{0,i}^{(s)}$ is the permutation operator acting on
$\CC^{2s+1}\otimes \CC^{2s+1}$. They satisfy also the famous
Yang-Baxter equation \cite{yang,baxter}:
\begin{equation}
\label{eq:YBE}
R^{(s_i,s_j)}_{i,j}(u-v)R^{(s_i,s_k)}_{i,k}(u-w)
R^{(s_j,s_k)}_{j,k}(v-w)
=R^{(s_j,s_k)}_{j,k}(v-w)R^{(s_i,s_k)}_{i,k}(u-w)
R^{(s_i,s_j)}_{i,j}(u-v) \;.
\end{equation}

Finally, we introduce the following transfer matrices
\begin{equation}
\label{def:transf}
t^{(\bar s_i)}(u)=\text{tr}_{0}T^{(\bar s_i)}_0(u)\,.
\end{equation}
They lead to conserved quantities
\begin{equation}
I_n^{(s)}=\frac{d^n}{du^n}\ln t^{(s)}(u)\Big|_{u=0}
\mb{with} s\in\cS\;.
\end{equation}
The Hamiltonian is usually chosen as any linear combinations of the 
following conserved charges
\begin{equation}
H^{(s)}=iI_1^{(s)}\;,
\end{equation}
where the factor $i$ allows us to obtain a Hermitian operator.
However, the explicit computation of this Hamiltonian for an
$L_0$-regular spin chain become more involved since $R^{(s,s')}(0)$ 
is not a permutation for $s \neq s'$. Even  locality of this 
operator is not obvious.
Fortunately, by introducing new transfer matrix, we may 
construct new framework where the usual constructions work even for a 
general spin chain.
We illustrate that in the following subsection by computing the 
momentum and the Hamiltonian.

\subsection{Momentum and Hamiltonian}

In the homogeneous case ($L_0=1$), the transfer matrix at vanishing 
spectral 
parameter 
provides the one-step shift operator and the momentum is given by its
logarithm. In the case of $L_0$-regular spin chain, the one-step 
shift operator
is not any more conserved.
However, it is obvious that the $L_0$-shift operator, $\cS_{L_0}$,
must be conserved.
To express this operator in terms of 
the transfer matrices 
(\ref{def:transf}), we introduce the following transfer matrix
\begin{equation} 
\label{eq:deftg}
{\boldsymbol{t}}({\boldsymbol{u}})=t^{(s_{1})}(u_1)\ 
t^{(s_{2})}(u_2)\dots
t^{(s_{L_0})}(u_{L_0})
\end{equation}
where $u_1,\dots,u_{L_0}$ are different spectral parameters. 
Obviously, it commutes with any other 
transfer matrix $t^{(s_{i})}(v)$ and it may be written as follows
\begin{equation}
{\boldsymbol{t}}({\boldsymbol{u}})=tr_{a_1,\dots,a_{L_0}}\prod_{0\leq 
p<L/L_0}^{\longrightarrow}
\cR_{(a_1,\dots,a_{L_0}),(1+pL_0,\dots,(p+1)L_0)}(\boldsymbol{u})
\end{equation}
where we have introduced 
\begin{equation}
\cR_{(a_1,\dots,a_{L_0}),(b_1,\dots,b_{L_0})}(\boldsymbol{u})=
\Big( R^{(s_1,s_1)}_{a_1,b_1}(u_1)\dots 
R^{(s_{L_0},s_1)}_{a_{L_0},b_1}(u_{L_0})\Big)
\dots
\Big(R^{(s_1,s_{L_0})}_{a_1,b_{L_0}}(u_1)\dots 
R^{(s_{L_0},s_{L_0})}_{a_{L_0},b_{L_0}}(u_{L_0})\Big)\;.
\end{equation}
The importance of this new operator lies in the fact that it is 
regular i.e.
\begin{equation}
\label{eq:genreg}
\cR_{(a_1,\dots,a_{L_0}),(b_1,\dots,b_{L_0})}(\boldsymbol{u})
\Big|_{\boldsymbol{u}=\boldsymbol{0}}=
P^{(s_1)}_{a_1,b_1}P^{(s_2)}_{a_2,b_2}\dots P^{(s_{L_0})}_{a_{L_0},b_{L_0}}\;.
\end{equation}
To prove this regularity, we have used the regularity of the R-matrix 
as well as 
the unitarity relation for the vanishing spectral parameter (see 
relations (\ref{eq:prfused})).
Using this property, it is a standard computation to show that
${\boldsymbol{t}}({\boldsymbol{0}})$ provides the $L_0$-step shift 
operator $\cS_{L_{0}}$.
We can deduce from this operator, the momentum operator
$\wh \fp$ defined as
\begin{equation}
{\boldsymbol{t}}({\boldsymbol{u}})
\Big|_{\boldsymbol{u}=\boldsymbol{0}}=\cS_{L_0}
=\exp(-iL_0\wh \fp)\;.
\end{equation}

It is easy to shown that the gradient of the transfer matrix 
${\boldsymbol{t}}({\boldsymbol{u}})$ allows us to obtain the general 
Hamiltonian.
Indeed, we get
\begin{equation}
H=i \boldsymbol{\alpha}\cdot \nabla \ln
{\boldsymbol{t}}({\boldsymbol{u}})\Big|_{\boldsymbol{u}=\boldsymbol{0}}
\mb{with} \boldsymbol{\alpha} = (\alpha_{1},\dots,\alpha_{{L_0}})
\mb{and}
\nabla=
\left(\begin{array}{c} 
\frac{\prt}{\prt u_{1}} \\ \vdots \\
\frac{\prt}{\prt u_{L_{0}}} \end{array}\right)
\end{equation}
where $\alpha_{1},\dots,\alpha_{{L_0}}$ are free parameters.
Using definition (\ref{eq:deftg}) of the transfer matrix 
${\boldsymbol{t}}({\boldsymbol{u}})$, we can show that 
$H$ is a linear combination of $H^{(s)}$
\begin{equation} 
\label{eq:Hgeneral}
H=\sum_{s \in \cS}\theta_{s} H^{(s)}\;,
\end{equation}
where $\theta_s=\sum_{j=1}^{L_0} \delta_{s,s_j}\alpha_j$. In the 
following, we will take $\theta_{s}>0$, in order to have a correct 
particle interpretation for our models\footnote{The ground state 
configuration is not unique when 
 one of the coefficients $\theta_{s}$ vanishes, see 
 \cite{dVMN} where the particular case of alternating spin chain is 
 studied. The case of negative coefficients have been studied 
in \cite{DorfMei}, still for an alternating spin chain.}.
Locality of this general Hamiltonian can be seen using the following explicit formula
\begin{equation}
H=i\sum_{p=1}^{L/L_0}
P_{1+(p-1)L_0,1+pL_0}\dots 
P_{pL_0,(p+1)L_0}\,
\boldsymbol{\alpha}\cdot\nabla\cR_{(1+(p-1)L_0,\dots,pL_0),(1+pL_0,\dots,(p+1)L_0)}(\boldsymbol{u})
\Big|_{\boldsymbol{u}=\boldsymbol{0}}\,.
\end{equation}

\section{Bethe ansatz\label{sec:bet}}

\subsection{Bethe equations}

To obtain the spectrum of the Hamiltonians $H$, we 
study,
as usual, the spectrum of the transfer matrix $t^{(s)}(u)$.
Its eigenvalues $\tau^{(s)}(\lambda)$ have
been computed by algebraic Bethe ansatz in
\cite{Kir,cas}
\begin{equation}
\label{eq:taus}
\tau^{(s)}(u)=\sum_{\alpha=0}^{2s}C_\alpha^{(s)}(u)
\prod_{p=1}^M\frac{(u-\lambda_p+i(s+1))(u-\lambda_p-is)}
{(u-\lambda_p+i(\alpha-s+1))(u-\lambda_p+i(\alpha-s))}
\end{equation}
where
\begin{equation}
C_\alpha^{(s)}(u)=\prod_{k=\alpha}^{2s-1}\prod_{s'\in \cS}
\left(\frac{u+i(k-s-s'+1)}{u+i(k-s+s'+1)}\right)^{L\de_{s'}}\,,\ 
\alpha<2s
\qmbox{and} C_{2s}^{(s)}(u)=1\,.
\end{equation}
The parameters $\{\lambda_n\}$ are the Bethe roots satisfying the 
Bethe equations:
\begin{equation}
\label{eq:bethe2}
\prod_{s\in\cS}
\left(\frac{\lambda_n+i s}{\lambda_n-i s}\right)^{L\de_{s}}
=
-\prod_{p=1}^M\frac{\lambda_n-\lambda_p+i}{\lambda_n-\lambda_p-i}
\qmbox{for}1\leq n \leq M\;.
\end{equation}
and $M$ is an integer depending on the choice of the eigenvectors.
These parameters are linked to the
total spin of the chain \cite{thermy}
\footnote{Be careful, there is a factor 2 between the spin $S$ 
defined here 
and the one defined in \cite{thermy}.}
\begin{equation} \\
\label{eq:Sto}
S=S_0-M\;.
\end{equation}
where $S_0=L\sum_{s\in\cS} s\de_{s}$ is the highest spin reached in 
this model.
Let us remark that the Bethe equations do not depend on the choice of 
the Hamiltonian.

The momentum $\fp$ (eigenvalues of $\wh 
\fp$) and 
the energies, $E^{(s)}$ (eigenvalues of the 
Hamiltonian $H^{(s)}$), are given by
\begin{eqnarray}
\fp&=&i\sum_{s\in\cS}\de_s\ \sum_{n=1}^M
\ln\left(\frac{\lambda_n+i s}{\lambda_n-i s}\right)\ 
\text{mod}(\frac{2\pi}{L_0})=
\sum_{s\in\cS}\de_s\ \sum_{n=1}^M
\left(\varphi_{2s}(\lambda_n)+\pi\right)\ 
\text{mod}(\frac{2\pi}{L_0})\\
\label{eq:nrj1}
E^{(s)}&=&-\sum_{k=1}^M\frac{2 s}{(\lambda_k)^2+ s^2}
\end{eqnarray}
Let us remark that each $\lambda_k$ provides a negative energy. Then, 
the state with
$M=0$,
which is the pseudo-vacuum used in the procedure of the algebraic 
Bethe ansatz, is the
state with highest energy.
We are in the case of an 'anti-ferromagnetic' spin chain. The true 
vacuum will be studied
in the following. Multiplying the Hamiltonian by a negative constant, 
we describe a
'ferromagnetic' spin chain.

\subsection{String hypothesis \label{sec:string}}

We want to study the previous models in the thermodynamical limit 
($L\rightarrow\infty$)
and, in particular, to compute the energy of the vacuum state as well 
as
the one of the first excited states.
In the thermodynamical limit, it is usual to use
the string hypothesis which states that all the Bethe roots
$\{\lambda_p\ , \ p=1,\dots,M\}$
gather into $\nu_m$ strings of length $2m$, called $2m$-strings, 
($m\in\half\ZZ_{>0}$)
of the following form
\begin{equation}
\lambda_{m,k}+i\,\alpha
\,,\quad \alpha=-m+\half,-m+\frac{3}{2},\dots,m-\half
\end{equation}
where $k=1,\dots,\nu_{m}$ and $\lambda_{m,k}$, the center of
the string, is real. We get
\begin{equation}
\label{eq:spin_string}
M=2\sum_{m\in\half\ZZ_{>0}} m\nu_m \qmbox{and} 
S=-2\sum_{m\in\half\ZZ_{>0}} m\nu_m+
L\sum_{ s\in\cS} \de_s\  s\;.
\end{equation}
\begin{rmk}\label{rmk:string}
In the string hypothesis, usually, we suppose also that the
finite size effects 
in the imaginary part are exponentially small in $L$. However, it is 
well-established that 
this assumption is wrong for spin $s$ chains with $s\geq 1$ (see, for 
example, the articles \cite{avdo,vewo} where this 
deviation has been computed numerically and analytically).
The general case treated here is certainly worst and, in general, the 
decay of the 
imaginary part will be of order $1/L$. However, to study the 
Bethe equations in this hypothesis is 
still interesting and fruitful. Indeed, the number of states obtained 
by this way is in agreement with the dimension of the Hilbert space 
(see section \ref{sec:com}), we can compute the energy of the 
antiferromagnetic vacuum and determine the dispersion relation for 
the 
first excited states.
\end{rmk}

Within this hypothesis, the Bethe equations (\ref{eq:bethe2}) can be 
transformed
and become equations in terms of the real centers of the strings 
only. After
taking the logarithm, we get, for $m\in \half \ZZ_{>0}$ and 
$k=1,\dots,\nu_m$,
\begin{equation}
\label{eq:bethe3}
-2\pi Q_{m,k}+L\sum_{s\in \cS} \de_s \Phi_{2 s}^{(m)}(\lambda_{m,k})=
\sum_{p\in \half \ZZ_{>0}}\sum_{\ell=1}^{\nu_p}\Phi_2^{(p,m)}
(\lambda_{m,k}-\lambda_{p,\ell})
\end{equation}
where $Q_{m,k}$ are half-integers and
\begin{eqnarray}
\Phi_2^{(p,m)}(\lambda)&=&
\displaystyle\vph_{2p+2m}(\lambda)+\vph_{2|p-m|}(\lambda)
+2\sum_{\alpha=|p-m|+1}^{p+m-1}\vph_{2\alpha}(\lambda)
\\
\label{eq:Phi}
 \Phi_{p}^{(m)}(\lambda) &=&
\sum_{\alpha=\vert\frac{p}{2}-m+\half\vert+1}^{\frac{p}{2}+m-\half}
\vph_{2\alpha}(\lambda)
+\theta(p>2m-1)\vph_{p-2m+1}(\lambda)
\,,\quad p\in\ZZ_{>0}\ ,\ m\in\half\ZZ_{>0}\;.\qquad
\end{eqnarray}
The numbers $Q_{m,k}$ are supposed to be quantum numbers i.e. 
for one set there exists 
one and only one solution to Bethe equations (\ref{eq:bethe3}).  
Constraints on these numbers will be given in  
section \ref{sec:com}. 

Within the string hypothesis, the momentum and the energies 
(\ref{eq:nrj1}) 
become, 
for $s\in \cS$,
\begin{eqnarray}
\fp&=&\sum_{s\in\cS}\de_s\sum_{m\in \half\ZZ_{>0}}\sum_{k=1}^{\nu_m}
\Phi_{2 s}^{(m)}(\lambda_{m,k})+2\pi \sum_{m\in \half\ZZ_{>0}}m\nu_m 
\quad
\text{mod}(\frac{2\pi}{L_0})\;,\\
\label{eq:nrj2}
E^{(s)}&=&- \sum_{m\in \half\ZZ_{>0}}\sum_{k=1}^{\nu_m}
\sum_{\alpha=-m+\half}^{m-\half}
\frac{2(\alpha+ s)}{(\lambda_{m,k})^2+(\alpha+ s)^2}
=- \sum_{m\in \half\ZZ_{>0}}\sum_{k=1}^{\nu_m}
\Psi_{2 s}^{(m)}(\lambda_{m,k})
\end{eqnarray}
where $\Psi_p^{(m)}(\lambda)$ is the derivative of 
$\Phi_p^{(m)}(\lambda)$.
We will need also to define $\Psi_2^{(p,m)}(\lambda)$, the 
derivative of
$\Phi_2^{(p,m)}(\lambda)$. The explicit form of these functions may be found in \cite{thermy}.

\subsection{Valence and completeness of Bethe states \label{sec:com}}

{From} equation (\ref{eq:bethe3}) we can get bounds on 
$Q_{m,k}$ \cite{thermy}:
\begin{equation}
Q_{m,max}=
\frac{1}{2}\Big(\nu_m-1+2\sum_{i=1}^{L}\min(m,s_i)
-4\sum_{n\in \frac{1}{2}\ZZ^+}\min(m,n)\nu_n\Big)
\mb{and} Q_{m,min}=-Q_{m,max}\,.
\end{equation}
Now we can define the valence, which is the number of allowed quantum 
numbers $Q_{m}$ for a given configuration $\{\nu\}$:
\begin{eqnarray}
P_m(\nu)&=&2Q_{m,max}+1=
2\sum_{j=1}^L \min(m,s_j)-4\sum_{n \in 
\frac{1}{2}\ZZ^+}\min(m,n)\nu_n +\nu_m
\label{val}
\end{eqnarray}
As explained previously, to each set of quantum numbers $Q$ 
corresponds one Bethe eigenstate. Then, 
to be sure that this method gives all the eigenstates, we must prove 
that the number of eigenstates obtained by Bethe ansatz is equal to 
the dimension of the starting Hilbert space.
Let us recall that the Bethe eigenvectors are highest weight for the 
$gl(2)$ symmetry, so that
a Bethe eigenvector is $(2S+1)$-degenerated, where $S=S_0-M$ is its 
total spin (\ref{eq:Sto}).
Thus, given the valence (\ref{val}), the number of the eigenstates 
for a given M obtained by Bethe ansatz is 
\begin{equation}
Z_M^{bethe}=(2S_0-2M+1)\sum_{\atopn{\{\nu_m\}}{2 \sum k\nu_k = M}} 
\prod_{m\in\half\ZZ_{+}}
\left(\begin{array}{c} P_m(\nu) \\
\nu_m \end{array} \right)
\label{numvectorM}
\end{equation}
where we sum over all the possible configurations $\{\nu\}$ (number of 
string of 
each type) and $\left(\begin{array}{c} a \\ c \end{array} \right)$ 
is the binomial coefficient.

Following the previous work \cite{KIRI} on this problem, we can 
compute explicitly this number.
The proof is based on the following combinatorial identity, for 
$\{b\}$ a set of real numbers and $\{\nu\}$
a set of positive integers,
\begin{eqnarray}
\sum_{M=0}^{\infty} Z(\{b\},M) x^M=(1-x)\prod_{n=1}^{\infty} 
(1-x^n)^{b_n} \label{eq:Z}
\end{eqnarray}
where we have introduced 
\begin{eqnarray}
Z(\{b\},M) &=&
\sum_{\atopn{\{\nu_m\}}{2 \sum_{k} k\nu_k = M}} 
\prod_{m\in\half\ZZ_{+}}
\left(\begin{array}{c} A_m(\nu,b) \\ \nu_m \end{array} \right)
\\
A_m(\nu,b) &=& -\sum_{j=1}^{2m} (2m-j+1)b_j-2M+4\sum_{n>m}(n-m)\nu_n 
+\nu_m
\label{combK}
\end{eqnarray}
For the following particular choice of the set $\{b\}$
\begin{eqnarray}
b_1&=&-L \\
b_m&=&\begin{cases}
0 \mb{,} m \neq 2\bar s_j+1 \\
{L}\de_j \mb{,} m = 2\bar s_j+1
\end{cases}\qquad m=2,3,\ldots
\end{eqnarray}
we have
$A_{m}(\nu,b)=P_{m}(\nu)$
and (\ref{eq:Z}) 
in the limit $x \to 1$ gives (see 
\cite{KIRI} for details):
\begin{equation}
\sum_{M=0}^{S_{0}}Z_M^{bethe}=
\prod_{j=1}^{\cL}(2\bar s_j+1)^{L \de_j}
\end{equation}
The L.H.S. is the total number of states we get from the Bethe 
equations in the string 
hypothesis, while the R.H.S. is the total dimension of the Hilbert
space.
Thus, the Bethe ansatz in the string hypothesis leads to a
complete basis of states.

\section{Vacuum state\label{sec:vac}}

For any choice of Hamiltonian $H^{(s)}$ (or any linear combination 
with positive coefficients), 
the contribution 
to the energy of any Bethe roots is negative
(see eq. (\ref{eq:nrj1}) or (\ref{eq:nrj2})). Then, to obtain the 
true ground
state (i.e. to minimize the energy), we look for a configuration with 
a maximum number 
of roots. So, it is natural to introduce
the vacuum state defined by
\begin{equation}
\label{eq:vac}
P_n(\nu)-\nu_n=0\qmbox{for} n\in \half\,\ZZ_{>0}\;.
\end{equation}
where the valences $P_n(\nu)$ have been defined by (\ref{val}).
This constraint has been solved in \cite{thermy} and one finds a 
unique configuration
characterized by
\begin{equation}
\label{eq:confvac}
\nu_{s}=
\begin{cases}
\displaystyle \frac{L\de_s}{2}\qquad& s\in \cS\\
0& \text{otherwise}
\end{cases}
\end{equation}
 One interprets it as $\cL$ filled Fermi seas of $2s$-string (for 
$s\in\cS$).
{From} now on, this state becomes the reference state.
As we will see in section \ref{sec:exct}, any excited states have an 
energy greater 
than the one of this reference state, whatever the Hamiltonian 
$H^{(s)}$ one considers:
the vacuum state is also its ground state.
It is non degenerate since its spin vanishes (which is easily deduced 
from 
(\ref{eq:spin_string})).

Relation (\ref{eq:vac}) implies that the quantum numbers fulfil all 
the possibilities: 
\begin{equation}
Q_{s,k}=k-\half-\half \nu_{s}
\qmbox{with}k=1,\dots,\nu_{s}
\mb{and}s\in \cS\;.
\end{equation}
For the vacuum state, in the thermodynamical limit,
the Bethe roots $\{\lambda_{s,k}\ |\ k=1,\dots,\nu_s, s\in\cS\}$ 
become dense in $\RR$
and we can replace them by their density $\sigma^{(0)}_{s}(\lambda)$.
Then, the Bethe equations (\ref{eq:bethe3})
can be transformed to the following integral equations, for $s\in\cS$,
\begin{equation}
\label{eq:bethe4}
-2\pi\sigma^{(0)}_{s}(\lambda_0)
+\sum_{r\in \cS} \de_{r} \Psi_{2 r}^{(s)}(\lambda_0)=
\sum_{r\in\cS} \int_{-\infty}^\infty d\lambda\  
\sigma^{(0)}_{r}(\lambda)\
\Psi_2^{(r,s)}(\lambda_0-\lambda)\;.
\end{equation}
Solving these integral equations, we get the densities \cite{thermy}
\begin{equation}
\label{eq:densvac}
\sigma^{(0)}_{s}(\lambda)=\frac{\de_s}{2}\ \frac{1}{\cosh(\pi 
\lambda)}
=\de_s \sigma^{(0)}(\lambda)\;.
\end{equation}
The computation of these densities allows us to determine the 
energies of the vacuum
\begin{theorem}\label{th:nrjvac}
The energies per site (energy densities), eigenvalues of the 
Hamiltonians $H^{(s)}$
($s\in\cS$) divided by the length $L$, for an $L_0$-regular spin chain are given by
\begin{equation}
\cE_0^{(s)}=-\sum_{s'\in\cS}\ \de_{s'} \
\left(\psi\Big(\frac{ s'+s+1}{2}\Big)-
\psi\Big(\frac{|s'-s|+1}{2}\Big)\right)
\end{equation}
where $\psi(x)$ is the Euler digamma functions.
\end{theorem}
\prf
Replacing the sum $\sum_{k=1}^{\nu_m}$ in (\ref{eq:nrj2})
by an integral, we get the energies for the vacuum
\begin{eqnarray}
E_0^{(s)}=-L\sum_{s'\in\cS} \de_{s'}\ \int_{-\infty}^\infty d\lambda
\ \sigma^{(0)}(\lambda)
\Psi_{2 s}^{(s')}(\lambda)
=-2\pi L \sum_{s'\in\cS} \de_{s'}\ \int_{-\infty}^\infty dp
\ \wh \sigma^{(0)}(p)
\wh \Psi_{2s}^{(s')}(p)\;.\label{eq:nrjvacc}
\end{eqnarray}
The second equality is obtained via the Plancherel's theorem.
Using the explicit forms of $\wh \Psi$ \cite{thermy}
and the one of $\wh \sigma^{(0)}$ (see \ref{eq:densvac}), we get
the result.
\finprf
Similarly, we can prove that, for the vacuum, the momentum is given by
\begin{eqnarray}
\fp_0
&=&\pi L \sum_{s\in \cS}s \de_s\quad
\text{mod}(\frac{2\pi}{L_0})\,.
\label{eq:vac0}
\end{eqnarray}

\section{Excited states \label{sec:exct}}


\subsection{Characterization of excited states \label{sec:qex}}

The excited states are
obtained by creating holes in the filled Fermi seas
of $2s$-strings ($s\in\cS$) or creating new $2r$-strings with
$r\in \cR$.
Such states are 
characterized by the following configuration:
\begin{equation}\label{eq:confe}
\wt\nu_{s}=\nu_{s}-\mu_{s}\mb{for} s\in\cS
\mb{and} \wt\nu_{r}\geq 0 \mb{for} r\in\cR \, ,
\end{equation}
where we kept the notation $\nu_{s}$ ($s\in \cS$) for the vacuum
configuration (\ref{eq:confvac}) while the positive integers 
$\wt\nu_{r}$ ($r\in\cR$) correspond to the 
numbers of new $2r$-strings with centers $\lambda_{r,\ell}$ 
($\ell=1,\dots,\wt\nu_{r}$). The
corresponding valences are given
by, for $n\in\half\ZZ_{>0}$,
\begin{eqnarray}
\label{eq:val-e}
\wt P_n(\wt \nu)&=&\wt \nu_n+
4\sum_{s\in\cS}\min(n,s)\,\mu_{s}-4\sum_{r\in\cR}\min(n,r)\,\wt\nu_{r}\;.
\end{eqnarray}

Since $\wt P_s$ depends on $\wt\nu$, $\mu_s$ is not the 
number of holes in the sea of $2s$-strings: this physical quantity is 
rather defined 
by
\begin{equation}
\cD_s=\wt P_{s}(\wt \nu)-\wt \nu_{s} \mb{for} s\in\cS\,.
\end{equation}
It is the number of unused values, $\wt Q_{s,d}$ ($d=1,\dots,\cD_s$), 
in the set
$\{\frac{1-\wt P_s(\wt \nu)}{2},\dots,\frac{\wt P_s(\wt \nu)-1}{2}\}$ 
of possible 
choices for the quantum numbers in the sea of $2s$-string.
We denote by $\cD=\sum_{s\in\cS}\cD_s$ the total number of holes.
In the same way that one  associates a unique Bethe root, 
$\lambda_{s,k}$, 
to each quantum number $Q_{s,k}$, we introduce
$\wt \lambda_{s,d}$ associated to $\wt Q_{s,d}$. These numbers $\wt 
\lambda_{s,d}$ 
can be interpreted as rapidities of holes.
Let us remark that $\cD_s$ ($s\in\cS$) is always even (see equation 
(\ref{eq:val-e})).
This means that a single excitation is composed of two holes. This 
behavior appears already 
in the usual homogeneous spin $\half$ spin chain \cite{faddeev}.
To simplify some formulas, we will use also shorter notations
$\cD_j=\cD_{\bar s_j}$, $\mu_j=\mu_{\bar s_j}$ and $\wt 
\lambda_{j,d}=\wt 
\lambda_{\bar 
s_j,d}$.

Let us remark that the numbers $\{\mu\}$ are determined by the 
$\cD$'s and $\wt \nu_r$'s 
which allows us to express the numbers of unused quantum numbers $\cA_r=\wt P_r(\wt 
\nu)-\wt \nu_r$ for the new strings by
\begin{eqnarray}
\label{eq:con7}
\cA_{r}
&=&\frac{r-\bar s_j}{\bar s_{j+1}-\bar s_j}\cD_{j+1}
+\frac{\bar s_{j+1}-r}{\bar s_{j+1}-\bar s_j}\cD_j
-4\sum_{m\in\cR_j}\frac{(\bar s_{j+1}-\max(m,r))(\min(m,r)-\bar s_j)}
{\bar s_{j+1}-\bar s_j}\,\wt\nu_{m}\,, \ r\in\cR_{j}\qquad
\end{eqnarray}
Let us remark that the numbers $\cA_r$ is always even 
(see relation (\ref{eq:val-e})).

Now,  the number of eigenstates for a given number 
of holes is given by
\begin{equation}
\label{eq:ZZ}
Z(\{\cD\})=\sum_{\{\wt 
\nu_r\}}(2S+1)\prod_{r\in\cR}\left(\begin{array}{c} \cA_r+\wt\nu_r\\
\wt \nu_r \end{array} \right)
\end{equation}
where we sum over all the sets $\{\wt \nu_r\in\ZZ_{\geq 0}|r\in\cR\}$ 
such that
\begin{equation}\label{eq:ineq2}
\cA_r=\wt P_r(\wt \nu)-\wt \nu_r\geq 0\mb{for} r\in\cR\;.
\end{equation}
Inequality (\ref{eq:ineq2}) translates the obvious fact the we cannot 
have more quantum numbers 
than allowed.  The factor $2S+1$ comes from the 
degeneracy due to the $gl(2)$ symmetry with the total spin rewritten 
as follows
\begin{equation}
\label{eq:spin3}
S=\frac{\cD_{\cL}}{2}-2\sum_{r\in \cR_{\cL}}(r-\bar 
s_\cL)\wt\nu_{r}\,.
\end{equation}
Then, to simplify relation (\ref{eq:ZZ}), we invert relation 
(\ref{eq:con7}) to get, for $r\in\cR_j$
\begin{equation}
\label{eq:Anu}
\wt\nu_r=\half(\cA_{r-\half}+\cA_{r+\half}-2\cA_{r})
\end{equation}
with the conventions $\cA_{s}=\cD_{s}$ (for $s\in\cS$) and $\cA_0=0$.
Therefore, 
\begin{equation}
 Z(\{\cD\})=\prod_{j=0}^\cL Z_j\qmbox{with}Z_j=
\sum_{\cA_{\bar s_j+\half},\dots,\cA_{\bar 
s_{j+1}-\half}\in 2\ZZ_{\geq 0}}\ \prod_{r\in\cR_j}
\left(\begin{array}{c}\half(\cA_{r-\half}+\cA_{r+\half})\\ 
\cA_r\end{array} \right)\,.
\end{equation}
Finally, one can conjecture that these numbers are equal to, for 
$0\leq j\leq \cL-1$,
\begin{equation}
\label{eq:nu3}
Z_j=\frac{2^{\cD_j+\cD_{j+1}}}{\bar s_{j+1}-\bar 
s_j+1}
\sum_{q=1}^{2\bar s_{j+1}-2\bar s_j+1}
\sin^2\left(\frac{q\pi}{2\bar s_{j+1}-2\bar s_j+2}\right)
\cos^{\cD_j+\cD_{j+1}}\left(\frac{q\pi}{2\bar s_{j+1}-2\bar 
s_j+2}\right)\;.
\end{equation}
We do not know a full analytical proof of this result, but we 
proved it
 for $\bar s_{j+1}-\bar s_j=1,\frac32,2,\ldots,\frac72$ 
by brute force calculations on 
binomial coefficients that we do not wish to reproduce here. 
Remark that a similar feature appears also in the counting of states 
for the homogeneous highest spin 
XXZ model studied in \cite{berk} (see also section \ref{sec:Sm}).
Finally, we can also obtain an exact closed form for $Z_\cL$ given by
\begin{equation}
\label{eq:degS}
Z_\cL=2^{\cD_\cL}\;.
\end{equation}

\subsection{Density of roots for excited states \label{sec:deex}}

Now, we are in position to compute the densities of the Bethe roots 
corresponding to the states 
defined in the previous subsection.
For the configuration (\ref{eq:confe}), the Bethe equations 
(\ref{eq:bethe3}) for
$m\in\cS$, in the thermodynamical limit, provides a linear integral 
equation for the
densities
$\sigma_{s}(\lambda)$ of $2s$-strings:
\begin{eqnarray}
\label{eq:bethegen}
&&\hspace{-2cm}-2\pi\Big[\sigma_{s}(\lambda_0)+\frac{1}{L}
\sum_{d=1}^{\cD_s} \delta(\lambda_0-\wt\lambda_{s,d})\Big]
+\sum_{s'\in \cS} \de_{s'}
\int_{-\infty}^{\infty} \Psi_{2 
s'}^{(s)}(\lambda)\sigma_{s'}(\lambda)d\lambda\nonumber\\
&&\hspace{2cm}=
\sum_{s'\in \cS}\int_{-\infty}^{\infty}
\Psi_2^{(s',s)}
(\lambda_0-\lambda)\sigma_{s}(\lambda)d\lambda +\frac{1}{L}
\sum_{r\in \cR}\sum_{\ell=1}^{\tilde \nu_r}
\Psi_2^{(r,s)}
(\lambda_0-\lambda_{r,\ell})
\end{eqnarray}
There are also other Bethe equations for $\lambda_{r,\ell}$, 
$r\in\cR$. We postpone 
their study in
section \ref{sec:link}.

These densities can be computed \cite{therm2}:
\begin{equation}
\label{eq:t}
\sigma_{s}(\lambda)=\sigma^{(0)}_{s}(\lambda)
+\frac{1}{L}\big(\fr_{s}(\lambda)+\fc_{s}(\lambda)\big)
\end{equation}
where $\sigma^{(0)}_{s}(\lambda)$ is the density 
(\ref{eq:densvac}) of the
vacuum, $\fr_s(\lambda)$ is the correction due to the holes
and $\fc_{s}(\lambda)$ is the polarization due to the new strings.
The explicit form of these corrections  
\cite{therm2} reads:
\begin{eqnarray}
 \fr_{\bar s_j}(\lambda)&=&
\frac{1}{2\pi}\left(
\sum_{d=1}^{\cD_{{j-1}}}
\kappa^{(\hbar_{j-1})}_{2(\bar s_j-\bar s_{j-1})-1}(\lambda-\wt 
\lambda_{{j-1},d})
+\sum_{d=1}^{\cD_{{j+1}}}
\kappa^{(\hbar_{j})}_{2(\bar s_{j+1}-\bar s_j)-1}(\lambda-\wt 
\lambda_{{j+1},d})
\right.\nonumber\\
&&\left. \hspace{2cm}
+\sum_{d=1}^{\cD_{{j}}}
\Big(
\kappa^{(\hbar_{j})}_1(\lambda-\wt \lambda_{{j},d})+
\kappa^{(\hbar_{j-1})}_1(\lambda-\wt 
\lambda_{{j},d})-2\pi\delta(\lambda-
\wt \lambda_{{j},d})\Big)
\right)
\quad
\label{eq:densite-correction}\\
\fc_{\bar s_j}(\lambda)&=&-\frac{1}{2\pi}
\sum_{m\in\cR_{{j-1}}}
\sum_{\ell=1}^{\wt \nu_m}
 \gamma^{(\hbar_{j-1})}_{2(\bar s_j-m)}(\lambda-\lambda_{m,\ell})
-\frac{1}{2\pi}\sum_{m\in\cR_{{j}}}
\sum_{\ell=1}^{\wt \nu_m}
 \gamma^{(\hbar_{j})}_{2(m-\bar s_j)}(\lambda-\lambda_{m,\ell})
\quad\label{eq:c}
\end{eqnarray}
where we have introduced $\displaystyle\hbar_j=\frac{\pi}{2(\bar 
s_{j+1}-\bar s_j)}$.
We set by convention $\cD_0=0=\cD_{\infty}$ (we recall the 
conventions $\bar s_0=0$ and $\bar s_{\cL+1}=\infty$).

\subsection{Energy and dispersion law \label{sec:nrj-dl}}

The densities given in previous section \ref{sec:deex} allow us to 
compute the
contribution at order $1/L$ of
the first excited states to the energies.
\begin{theorem}
The energy densities at order $1/L$ for the configuration 
(\ref{eq:confe}) are
$\cE^{(s)}=\cE^{(s)}_0+\frac{1}{L}\Delta E^{(s)}$
(for $s\in\cS$)
with $\cE^{(s)}_0$ given in theorem \ref{th:nrjvac} and
\begin{equation}
\Delta E^{(s)}=\sum_{d=1}^{\cD_s}\frac{\pi}
{\cosh\big(\pi \wt\lambda_{s,d}\big)}\;.
\end{equation}
\end{theorem}
\prf
There are three contributions to  the energies due to
$\fr$, $\fc$ and $\lambda_{m,\ell}$ (with $m\notin \cS$ and $s\in \cS$):
\begin{eqnarray}
\Delta E^{(s)}&=&-2\pi \sum_{s'\in\cS} \ 
\int_{-\infty}^\infty dp
\ (\wh\fr_{s'}(p)+\wh\fc_{s'}(p))
\wh\Psi_{2s}^{( s')}(p)-
\sum_{r\in\cR}\sum_{k=1}^{\wt \nu_r}
\Psi_{2s}^{(r)}(\lambda_{r,k})\;.
\end{eqnarray}
Using the explicit forms (\ref{eq:densite-correction}) and (\ref{eq:c}) of $\wh \fr$ and $\fc$, 
we prove the theorem.
\finprf
Let us emphasize the remarkable simplicity of this result although we 
deal with any
$L_0$-regular $gl(2)$ spin chain. We remark that the contribution to 
the energy
$E^{(\bar s_j)}$
(eigenvalues of the Hamiltonian constructed from the monodromy matrix 
with the
auxiliary space in the spin $\bar s_j$) involves only the holes in 
the sea of strings
of length $2\bar s_j$. The holes in the other seas as well as the new 
strings have a
vanishing energy.


Similarly, we can compute the eigenvalues of the impulsion and we 
obtain
\begin{equation}
\label{eq:mom}
\fp
=\fp_0+\sum_{s\in\cS}\ \de_s\ \sum_{d=1}^{\cD_s}
\Big(\arctan(\sinh(\pi \wt \lambda_{s,d})) +\frac{\pi}{2} \Big)
=\fp_0+
\sum_{s\in \cS}\ \sum_{d=1}^{\cD_s} \fp^{(s)}(\wt \lambda_{s,d})\;,
\end{equation}
where we have introduced
\begin{equation}
\label{eq:mom2}
\fp^{(s)}(\lambda)=\de_s\arctan(\sinh(\pi 
\lambda))+\frac{\de_s\pi}{2}\; .
\end{equation}
We recall that $\fp_0$ is the momentum of the vacuum defined by 
(\ref{eq:vac0}). 
Let us remark that $0<\fp^{(s)}<\de_s\pi$.
Then, we can deduce the dispersion law for these excited states, for 
$s\in\cS$,
\begin{equation}
 \Delta E^{(s)}=\pi \sum_{d=1}^{\cD_s}\
\sin\Big(\frac{\fp^{(s)}(\wt \lambda_{s,d})}{\de_s}\Big)\;.
\end{equation}
Thus, we conclude that, for the Hamiltonian $H^{(s)}$, the speed of 
sound\footnote{We remind
that the speed of sound is defined as the derivative of the energy
w.r.t. the momentum at the Fermi surface.} of the holes in the filled 
seas of
$2s$-string is equal to $\pi/\rho_s$ whereas it is $0$ for the holes 
in the
seas of $2s'$-string ($s'\in\cS$ and $s'\neq s$).

Choosing $\theta_s=\de_s$, we get the Hamiltonian 
with the energy 
$\Delta E=\sum_{s\in\cS}\de_s \Delta E^{(s)}$.
In this case, all the holes have the same speed of sound 
($=\pi$). Then, we deduce that this Hamiltonian
is a good candidate 
to be described by a continuum model which is conformal 
(see e.g. \cite{dewo,dev}).

\subsection{Interpretation in terms of particles \label{sec:cou}}

For the general Hamiltonian (\ref{eq:Hgeneral}), 
the excited states, characterized by $\cD_{s}$ holes 
in the seas of $2s$-string, can be interpreted as
$\cD_s$ particles like excitations with the dispersion 
law $E(p)=\pi \theta_s\sin(p/\de_s)$. 
We will call them particle of type $j$ when they correspond to a hole in the sea of 
$2\bar s_j$-string ($j=1,2,\dots,\cL$).

As explained at the end of section \ref{sec:qex}, for a given number 
of holes,
the state is degenerated due to the different possibility for the 
numbers 
$\wt \nu_r$ ($r\in \cR$) of new strings. In terms of particles, this 
degeneracy
is interpreted as a presence of an internal degree of freedom for the 
particles.

To understand degeneracy ($\ref{eq:degS}$), we associate to each 
particle
of type $\cL$ a spin $\half$ under the $gl(2)$ symmetry algebra.
This gives a space of dimension $2^{\cD_\cL}$ which is in agreement 
with ($\ref{eq:degS}$).

\begin{figure}[htb]
\begin{center}
\begin{minipage}[h]{12cm}
\epsfig{file=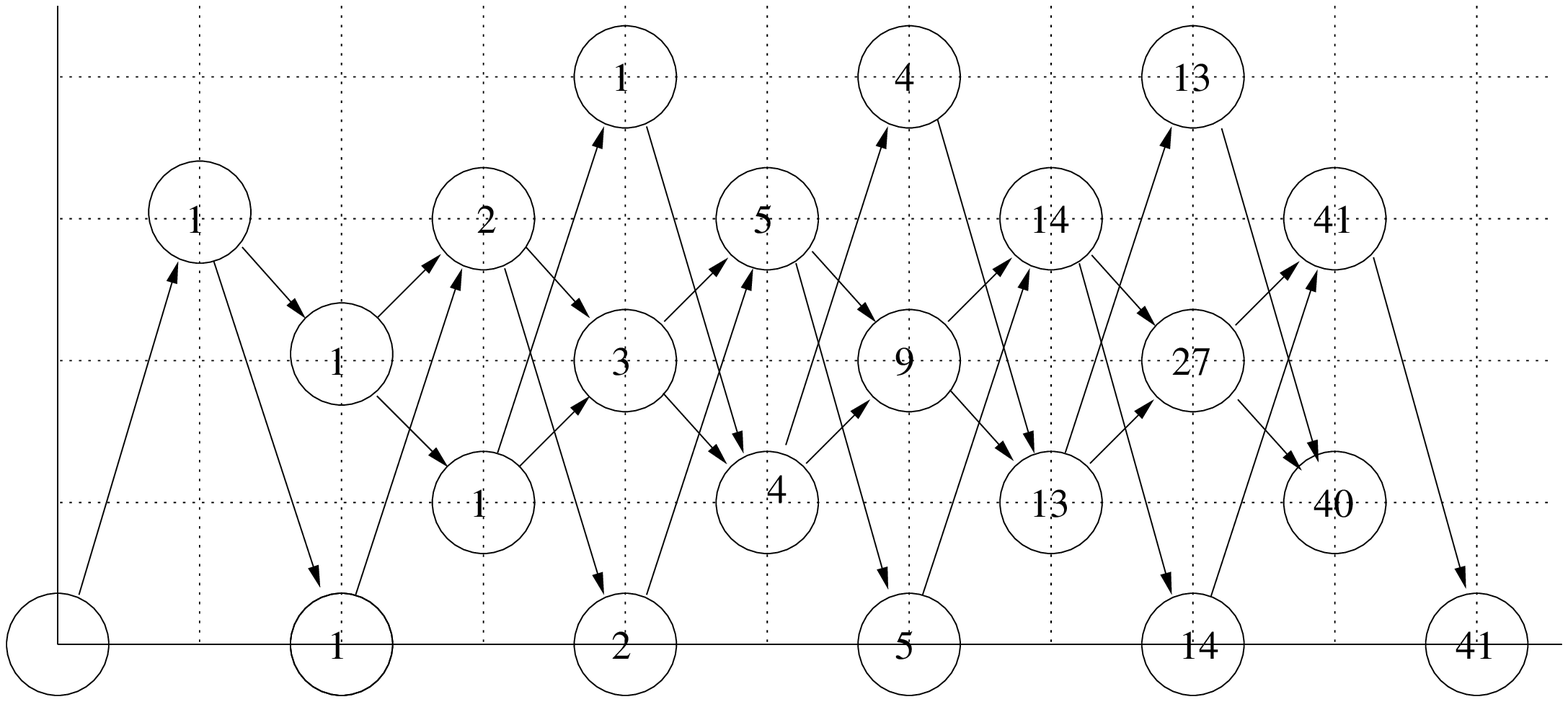,height=5cm,width=12cm}
\caption{
Path corresponding to $\cH^{RSOS}(10;0;2)$.
We have indicated the number of paths arriving to each 
allowed point.
\label{fig:rsoss2L8j3}}
\end{minipage}
\end{center}
\end{figure}

\begin{figure}[htb]
\begin{center}
\begin{minipage}[h]{12cm}
\epsfig{file=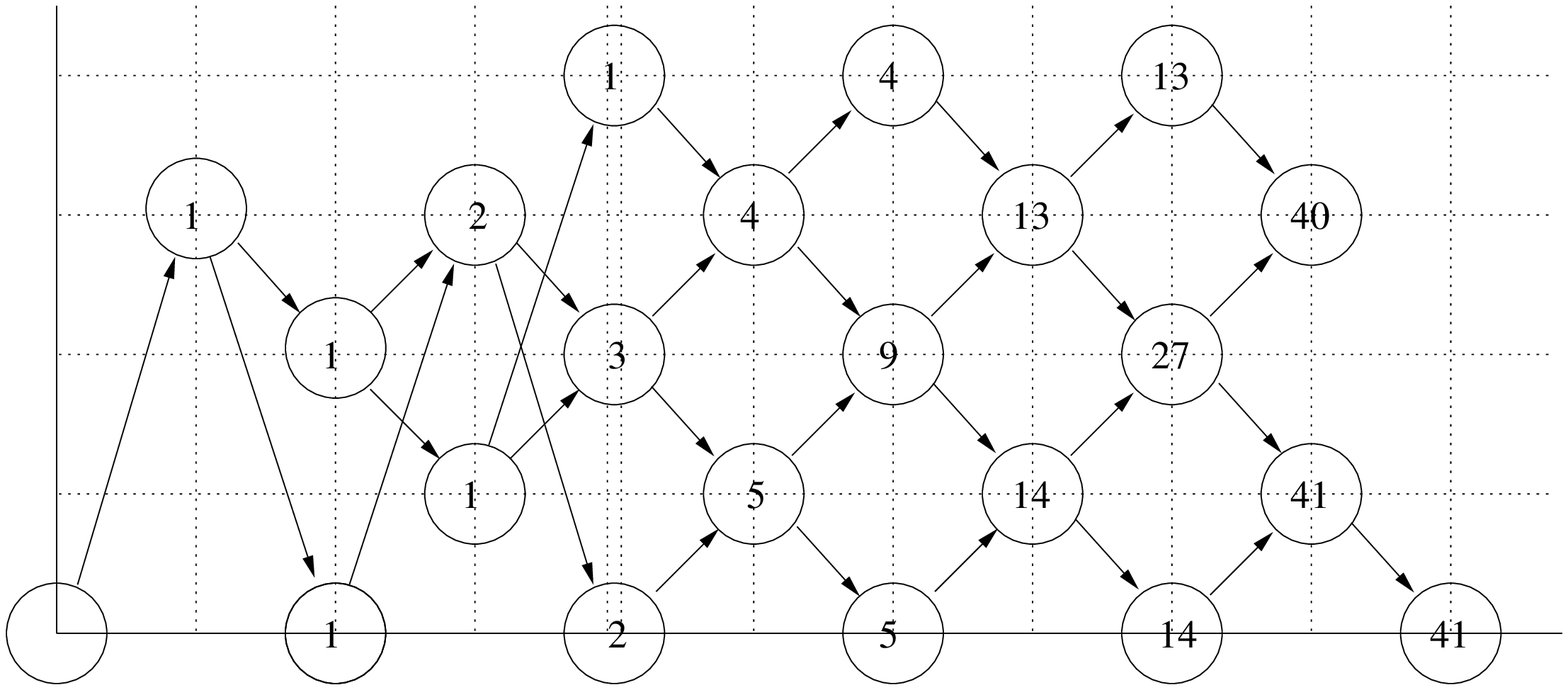,height=5cm}
\caption{Path corresponding to $\cH^{RSOS}(4;6;2)$. 
 The double dashed line indicates the changes in the height of the 
jumps.
\label{fig:rsoss2L8j1}}
\end{minipage}
\end{center}
\end{figure}

The particles of type $j$ ($j\neq\cL$) are scalar under the $gl(2)$ 
symmetry algebra (see eq. (\ref{eq:spin3}): $\cD_j$, 
$j\neq\cL$, does not 
appear in the 
expression of the spin). However, to explain the degeneracy given by 
($\ref{eq:nu3}$), we need
to introduce new internal degrees of freedom for the particles. The 
same problem occurs 
already in the case of the homogeneous spin chain and has been solved 
in 
\cite{NRe}. 
Generalizing this interpretation, we conjecture that the space with 
$\cD_1$ 
particles of
type $1$, $\cD_2$ particles of type $2$,...,$\cD_\cL$ particles of 
type $\cL$ 
is isomorphic 
to
\begin{equation}
\bigotimes_{q=1}^{\cL}\Big( \cH^{RSOS}(\cD_{q-1};\cD_{q};\bar 
s_q-\bar s_{q-1})\Big)\otimes(\CC^2)^{\otimes \cD_\cL}
\end{equation}
where $\cH^{RSOS}(\cD;\cD';\bar s)$ is the space of the integer 
sequences 
$(a_0,a_1,\dots,a_{\cD};b_0,b_1,\dots,b_{\cD'})$ with 
\begin{eqnarray}
&0\leq a_i\leq 2\bar s &\mb{;} 0\leq b_j\leq 2\bar s 
\label{eq:cons_rsos1}\\
&\displaystyle\frac{a_{i+1}-a_i+2\bar s-1}{2} \in \{0,1,\dots,2\bar s-1\} 
&\mb{;} 
\frac{b_{j+1}-b_j+1}{2} \in \{0,1\}\label{eq:cons_rsos2}\\
&2\bar s-2\leq a_j+a_{j+1}\leq 2\bar s+2\label{eq:cons_rsos3}
\end{eqnarray}
and the boundary conditions
$a_0=0$, $a_{\cD}=b_0$ and $b_{\cD'}=0$. In words, this space 
corresponds to a generalized RSOS model
\cite{bare} with a restriction parameter given by $2\bar s+2$ and for 
which the $\cD$ 
first sites have a jump of $2\bar s-1$ whereas the $\cD'$ last sites 
have a jump of 1. We give 2 examples in Figures \ref{fig:rsoss2L8j3} 
and \ref{fig:rsoss2L8j1}
of paths corresponding to integer sequences for the 
restriction parameter $2$ as well as the number of such paths.

We show below that this interpretation is in agreement with the 
dimension of the spaces under consideration. We will show in section 
\ref{sec:Smat} that it is also compatible with the structure of the 
scattering matrix.
\begin{proposition}
When $\cD$ and $\cD'$ are even, the number of states in 
$\cH^{RSOS}(\cD;\cD';\bar s)$ is given by
\begin{equation}
\frac{2^{\cD+\cD'}}{\bar s+1}
\sum_{q=1}^{2\bar s+1}
\sin^2\left(\frac{q\pi}{2\bar s+2}\right)
\cos^{\cD+\cD'}\left(\frac{q\pi}{2\bar s+2}\right)\;.
\end{equation}
\end{proposition}
\prf
The computation for $\cH^{RSOS}(0;\cD';\bar s)$ has been done in 
\cite{NRe} (see also 
\cite{berk}). Then, the proposition is demonstrated by remarking that 
the iterated relations
satisfied by the number of paths at the even sites are identical for 
the jump $1$ 
or jump $2\bar s -1$.
\finprf 
Let us stress that this number depends only on the sum $\cD+\cD'$. 
This point is illustrated 
by figures \ref{fig:rsoss2L8j3} and \ref{fig:rsoss2L8j1}
 where we can see that the numbers
of paths at the even sites are the same in both figures.

The dimension of the spaces $\cH^{RSOS}(\cD;\cD';\bar s)$ with 
$\cD+\cD'=\cD_j+\cD_{j+1}$ is in agreement with the degeneracy 
$Z_j$ (see (\ref{eq:nu3})).
In the following, from the study of the S-matrix, we will show that
the relevant models correspond to spaces $\cH^{RSOS}(\cD;\cD';\bar s)$
with $\cD=\cD_j$, $\cD'=\cD_{j+1}$ and $\bar s=\bar s_{j+1}-\bar 
s_{j}$.

Finally, we introduce a basis for this space given by
\begin{equation}\label{eq:basrsos}
E(\{a\}_{\cD};\{b\}_{\cD'})=E(a_0,a_1,\dots,a_{\cD};b_0,b_1,\dots,b_{\cD'})
\end{equation}
where the sets of integers $\{a_i\}$ and $\{b_i\}$ satisfy 
constraints (\ref{eq:cons_rsos1})-(\ref{eq:cons_rsos3}).

\section{S-matrix\label{sec:Smat}}

In this section, we want to compute the scattering matrix between the
particles we introduced above. 
To do that, we follow the construction done in \cite{korepin,KoIzBo}:
one considers $\cD$ particles on a circle with a large (but finite) 
circumference $L$ and computes 
the phase shift collected by one particle when it passes through all 
the other ones. Let us remark that, although the S-matrix is a bulk property,
we must consider a finite system to compute it by this method.

We will get an
expression (see equation (\ref{eq:ST})) which
depends explicitly on the new Bethe roots $\lambda_{r,k}$ ($r\in\cR$ 
and 
$k=1,\dots,\wt\nu_r$).
These new Bethe roots are not free: they are determined by Bethe 
equations 
(\ref{eq:bethe3}) for $m\in\cR$ which were not used up to now (see 
section
\ref{sec:link}). Unfortunately, these equations cannot be solved in 
general.
However, remarking that the equations we get are very similar to the ones 
of the 
homogeneous spin studied in \cite{NRe}, we will conjecture an 
explicit form for the scattering matrix.

\subsection{$\cD$-body scattering matrix \label{sec:Sg}}

In general,
for a given excited state with $\cD$ holes,
the scattering matrix, $S_{s,d}$ ($s\in\cS$ and $1\leq d \leq \cD_s$), between one particle with rapidity
$\wt \lambda_{s,d}$ and all 
the other particles is defined
by the following finite volume quantization of momentum of this particle:
\begin{equation}
\exp(i\fp^{(s)}(\wt \lambda_{s,d})L)\; S_{s,d}=1\;.
\end{equation}
Taking the logarithm of the previous relation, we get
\begin{equation}
\label{eq:quantif}
\fp^{(s)}(\wt \lambda_{s,d})+\frac{1}{L}\Phi_{s,d}=\frac{2\pi \wt 
Q_{s,d}}{L}
\end{equation}
where $S_{s,d}=\exp(i\Phi_{s,d})$ and $\wt Q_{s,d}$ are the lacking 
quantum number corresponding to $\wt \lambda_{s,d}$.

To find explicitly the form of $\Phi$, let us remark that,
using the definition of the density $\sigma$ and its explicit
form (\ref{eq:t}), we get
\begin{eqnarray}
\frac{2\pi\wt Q_{s,d}}{L}&=&
2\pi\int_{-\infty}^{\wt \lambda_{s,d}}\ d\lambda\ \Big[
\sigma_{s}(\lambda)
+\frac{1}{L}\sum_{q=1}^{\cD_s}\delta(\lambda-\wt \lambda_{s,q})
\Big]\;.
\end{eqnarray}
Then, we can show, due to the explicit form of the density 
$\sigma^{(0)}$
(see (\ref{eq:densvac})) and of the momentum (see (\ref{eq:mom2})), 
that
\begin{equation}
\fp^{(s)}(\lambda)=
2\pi \int_{-\infty}^{\lambda}\ d\mu\  \sigma^{(0)}_{ s}(\mu)\;.
\end{equation}
Finally, we deduce from (\ref{eq:quantif}), using (\ref{eq:t}) to express 
$\sigma_s(\lambda)$, that
\begin{equation}
\label{eq:phi}
\Phi_{s,d}=2\pi \int_{-\infty}^{\wt \lambda_{ s,d}}\ 
d\lambda\
\big[\ \fr_{ s}(\lambda)
+\sum_{q=1}^{\cD_s}\delta(\lambda-\wt \lambda_{s,q})\
+\fc_{ s}(\lambda)\big]
\;.
\end{equation}
Now, we use the Fourier transform of the densities 
(\ref{eq:densite-correction}) and (\ref{eq:c}) to express 
the Fourier transform of the derivative (w.r.t. $\wt\lambda_{s,d}$)
of $\Phi_{s,d}$ in terms of the functions $\wh \gamma$ and $\wh 
\kappa$.
Then, by using the results gather in section \ref{sec:not}, we get
the explicit form of the S-matrix of the $d^{ \text{th}}$ particle of 
type $j$ up to a multiplicative constant $C_j$:
\begin{equation} 
\label{eq:ST}
S_{j,d}=C_j\ \bigchk{S}_{j,d}(\wt \lambda_{\bar s_j,d})\ \wt 
S_{j,d}(\wt \lambda_{\bar s_j,d})
\end{equation}
where 
\begin{equation}
\label{eq:Sc}
\bigchk{S}_{j,d}(\lambda)=
\prod_{ m\in\cR_{{j-1}} }
\prod_{\ell=1}^{\wt \nu_{m}}
 G^{(\hbar_{j-1})}_{2(\bar s_j-m)}
(\lambda-\lambda_{m,\ell})
\prod_{ m\in\cR_{{j}} }
\prod_{\ell=1}^{\wt \nu_{m}}
G^{(\hbar_{j})}_{2(m-\bar s_j)}(\lambda-\lambda_{m,\ell})\quad\ 
\end{equation}
and
\begin{equation}
\wt S_{j,d}(\lambda)=
\prod_{q=1}^{\cD_{{j-1}}}
\cK^{(\hbar_{j-1})}_{\frac{\pi}{\hbar_{j-1}}-1}
(\wt \lambda_{j-1,q}-\lambda)
\prod_{q=1}^{\cD_{j}}
\cK_{1}^{(\hbar_{j-1})}
(\wt \lambda_{j,q}- \lambda)
\cK_{1}^{(\hbar_j)}
(\wt \lambda_{j,q}- \lambda)
\prod_{q=1}^{\cD_{{j+1}}}
\cK^{(\hbar_{j})}_{\frac{\pi}{\hbar_{j}}-1}
(\wt \lambda_{j+1,q}- \lambda)
\end{equation}
The constant may be computed by 
\begin{equation}
C_j=\lim_{\lambda\rightarrow-\infty}(\bigchk{S}_{\bar s_j,d}(\lambda)
\wt S_{\bar s_j,d}(\lambda))^{-1}\;,
\end{equation}
and, by using (\ref{eq:limits}) and knowing that 
$\exp(i\pi\cD_j)=1$ (since $\cD_j$ is even), we get
$C_j=\exp(-i\pi\mu_{\bar s_j})$.
Let us recall that $\mu_{\bar s_j}$, defined in section 
\ref{sec:qex}, is integer. Then, the constant $C_j$ is a sign.

We recall also that the
Bethe roots $\lambda_{r,\ell}$ (for $r\in \cR$)
are functions of $\wt \lambda_{\bar s_j,d}$
via the Bethe equations (\ref{eq:constraint}). Unfortunately, these
equations are not solved
explicitly
in general, so that (\ref{eq:Sc}) for $\lambda=\wt \lambda_{\bar 
s_j,d}$
cannot be brought explicitly to a 
factorized form depending on $\wt \lambda_{\bar s_k,q}-\wt 
\lambda_{\bar s_j,d}$ 
solely.

\begin{rmk} The previous form of the scattering matrix, computed within the string hypothesis, 
 proves that, 
for excitations with only holes and no new
string, the scattering matrix factorizes (as expected since we study 
integrable systems).
This factor is called usually the CDD factor of the scattering matrix.
When we have also new strings ($ \bigchk{S}\neq 1$), the 
factorization should also occur but we could not prove it on the 
general explicit form (\ref{eq:Sc}).
Indeed, one cannot (without solving (\ref{eq:constraint})) write the 
scattering matrix as a product
of functions of differences between hole rapidities.
Nevertheless, the factorization of the scattering matrix is assumed 
since the system is integrable.
\end{rmk}

\subsection{Bethe equations between holes and new strings 
\label{sec:link}}

There exist additional relations between the holes in the
 seas and the new strings provided by the Bethe equations 
(\ref{eq:bethe3}) for 
$m\in \cR$.
 They are given explicitly by, for
$1\leq j\leq \cL$ and $m\in\cR_{{j}}$,
\begin{eqnarray}
\label{eq:constraint}
-2\pi\,Q_{m,k}+
\sum_{d=1}^{\cD_{j}}
\Gamma^{(\hbar_j)}_{2(m-\bar s_j)}(\lambda_{m,k}-\wt\lambda_{\bar 
s_j,d})
+\sum_{d=1}^{\cD_{{j+1}}}
\Gamma^{(\hbar_j)}_{2(\bar s_{j+1}-m)}(\lambda_{m,k}-\wt\lambda_{\bar 
s_{j+1},d})
=\sum_{r\in\cR_{{j}}}
\sum_{\ell=1}^{\wt\nu_{r}}
F^{(r,m)}_{2}(\lambda_{m,k}-\lambda_{r,\ell})
\nonu
\end{eqnarray}
where $\hbar_j=\frac{\pi}{2(\bar s_{j+1}-\bar s_j)}$
and, when $m,r\in \cR_{{j}}$
\begin{eqnarray}
\label{eq:F2}
F_{2}^{(r,m)}(\lambda)&=&
\begin{cases} \displaystyle
\Gamma^{(\hbar_j)}_{4m-4\bar s_j}(\lambda)\,
+2\,\sum_{q=1}^{2m-2\bar s_j-1} \Gamma^{(\hbar_j)}_{2q}(\lambda)\,
 &\mbox{ if }m= r\\[1.2ex]
\displaystyle \Gamma^{(\hbar_j)}_{2r+2m-4\bar s_j}(\lambda)\,+
\,\Gamma^{(\hbar_j)}_{2\vert r-m\vert}(\lambda)\,
+2\sum_{q=\vert r-m\vert+1}^{r+m-2\bar s_j-1}
\Gamma^{(\hbar_j)}_{2q}(\lambda)
&\mbox{ if }m\neq r\end{cases}
\end{eqnarray}

Equation (\ref{eq:constraint}), for $m>\bar s_\cL$, may be rewritten 
as follows
(we used the convention $\bar s_{\cL+1}=\infty$)
\begin{equation}
\label{eq:constL}
-2\pi Q_{\bar m+\bar s_\cL,k}+\sum_{d=1}^{\cD_\cL}\Phi_{1}^{(\bar m)}
(\lambda_{\bar m+\bar s_\cL,k}-\wt \lambda_{\cL,d})
=\sum_{r\in \half \ZZ_{>0}}\sum_{\ell=1}^{\wt \nu_{r+\bar s_\cL,k}}
\Phi_2^{(r,\bar m)}(\lambda_{\bar m+\bar s_\cL,k}-\lambda_{r+\bar 
s_\cL,\ell})
\end{equation}
for $\bar m=1/2,1,3/2,\dots$. 
\begin{rmk}
Comparing with (\ref{eq:bethe3}),
we deduce that the Bethe roots $\lambda_{\bar m+\bar s_\cL}$ for the 
new strings 
(of length strictly
greater than $2\bar s_\cL$)
satisfy the Bethe equation of the center of $2m$-strings for
an auxiliary spin chain. More precisely, this auxiliary spin chain 
is an homogeneous spin $\half$ chain
with $\cD_\cL$ sites and inhomogeneity parameter $\wt 
\lambda_{\cL,d}$ at site $d$.
\end{rmk}

This remark, together with the degeneracy of the states $Z_\cL$
(see (\ref{eq:degS})), lead us to interpret the factors in (\ref{eq:ST}) 
containing $\hbar_{\cL}$, which are proportional to
\begin{equation}\label{eq:Smm}
\prod_{q=1}^{\cD_{\cL}}\cK_1^{(0)}(\widetilde \lambda_{\cL,q}-\wt 
\lambda_{\cL,j}) 
\prod_{m\in\{1/2,1,3/2,\dots\}}
\prod_{\ell=1}^{\wt \nu_{m+\bar s_{\cL}}}
e_{2m}(\wt \lambda_{\cL,j}-\lambda_{m+\bar s_{\cL},\ell})\,,
\end{equation}
as eigenvalues of the transfer matrix of an auxiliary XXX spin 
$\half$ chain with the spectral parameter taken at $\wt 
\lambda_{\cL,j}$. In fact, all the argumentation
do NOT depend on the value of $\bar s_\cL$, nor on all the other 
values present 
in the spin chain ($\bar s_1,\dots,\bar s_{\cL-1}$).
Therefore, the conclusions obtained for these factors of 
the S-matrix are completely similar to the usual case of the 
homogeneous spin $\half$ chain treated in \cite{faddeev}.
We come back on this point in the following section \ref{sec:Sm}.
\\

As for the homogeneous spin chain with spin greater than 1, new 
features appear due to
the presence of strings
of length smaller than $2\bar s_\cL$. To study the influence of these 
strings, we must
study
relation (\ref{eq:constraint}), for
$m\in\cR_{j}$ ($0\leq j<\cL$). Replacing the indices $m$ by 
$\bar s_{j+1}-m$, we get
an equivalent relation, for $m=\half,1,\dots,\bar s_{j+1}-\bar 
s_{j}-\half$ and for $k=1,2,\dots, \tilde \nu_{\bar s_{j+1}-m}$
\begin{eqnarray}
\label{eq:constraint3}
-2\pi\,Q_{\bar s_{j+1}-m,k}&+&
\sum_{d=1}^{\cD_{j}}
\Gamma^{(\hbar_j)}_{2\bar s_{j+1}-2\bar 
s_{j}-2m}(\lambda_{\bar s_{j+1}-m,k}-\wt\lambda_{\bar s_j,d})
+\sum_{d=1}^{\cD_{{j+1}}}
\Gamma^{(\hbar_j)}_{2m}(\lambda_{\bar s_{j+1}-m,k}
-\wt\lambda_{\bar s_{j+1},d}
)\nonu
&=&\sum_{r\in\{\half,1,\dots,\bar s_{j+1}-\bar s_{j}-\half\}}
\sum_{\ell=1}^{\wt\nu_{\bar s_{j+1}-r}}
F^{(\bar s_{j+1}-r,\bar s_{j+1}-m)}_{2}(\lambda_{\bar 
s_{j+1}-m,k}-\lambda_{\bar s_{j+1}-r,\ell})
\end{eqnarray}

\begin{rmk} \label{rm:xxz}
One recognizes in this relation the Bethe equations within the string 
hypothesis for an XXZ spin chain
with the deformation parameter at root of unity,
$q=e^{i\hbar_{j}}=\exp(i\frac{\pi}{2(\bar s_{j+1}-\bar s_j)})$, 
 and two types of spins: $\frac{2\bar s_{j+1}-2\bar 
s_j-1}{2}$ and $\half$.
Indeed, considering an XXZ spin chain with $\cD_{j}$ 
sites of spin $\frac{2\bar s_{j+1}-2\bar s_j-1}{2}$ 
and inhomogeneity parameters $\wt\lambda_{j,d}$ together with 
$\cD_{j+1}$ sites of spin $\half$ and inhomogeneity parameters 
$\wt\lambda_{j+1,d}$, one is led to the following Bethe equations
\begin{eqnarray} 
\label{eq:cBE2}
&&\prod_{d=1}^{\cD_j}
\frac{
\sinh(\hbar_j(x_m-\wt\lambda_{j,d}+i\frac{2\bar s_{j+1}-2\bar 
s_j-1}{2}))}
{\sinh(\hbar_j(x_m-\wt\lambda_{j,d}-i\frac{2\bar s_{j+1}-2\bar 
s_j-1}{2}))}
\prod_{d=1}^{\cD_{j+1}}
\frac{
\sinh(\hbar_j(x_m-\wt\lambda_{j+1,d}+\frac{i}{2}))}
{\sinh(\hbar_j(x_m-\wt\lambda_{j+1,d}-\frac{i}{2}))}\nonu
&&=\prod_{\ell=1,\ell\neq m}^{M}
\frac{
\sinh(\hbar_j(x_m-x_\ell+i))}
{\sinh(\hbar_j(x_m-x_\ell-i))}
\label{eq:bq}
\end{eqnarray}
As it has been discussed in \cite{taka,TakSuz}, the string content of 
the XXZ chain 
with $q$ root of the unity in the 
$\cD_j,\cD_{j+1}\to\infty$ 
limit consists in $2r$-strings with the 
restriction $2r<2(\bar s_{j+1}-\bar s_j)$ and roots with imaginary 
part 
$\bar s_{j+1}-\bar s_j$.
To identify (\ref{eq:constraint3}) with the Bethe equations for the 
center of 
2r-strings, one must consider $\lambda_{\bar s_{j+1}-r,k}$ has the 
center of the 2r-strings and
discard roots with imaginary part. We will call it a restricted 
string hypothesis.
\end{rmk}

There are two different S-matrices which depend on the parameters 
$\hbar_j$: 
$S_{j+1,d}$ and $S_{j,d}$. The factor in $S_{j+1,d}$ is proportional 
to
\begin{equation}
\label{eq:Sc1}
\prod_{q=1}^{\cD_{{j}}}
\cK^{(\hbar_{j})}_{\frac{\pi}{\hbar_{j}}-1}
(\wt \lambda_{j,q}-\wt \lambda_{j+1,d})
\prod_{q=1}^{\cD_{j+1}}
\cK_{1}^{(\hbar_{j})}
(\wt \lambda_{j+1,q}-\wt \lambda_{j+1,d})
\prod_{ m\in\cR_{{j}} }
\prod_{\ell=1}^{\wt \nu_{m}}
 G^{(\hbar_{j})}_{2(\bar s_{j+1}-m)}
(\wt \lambda_{j+1,d}-\lambda_{m,\ell})
\end{equation}
This value is similar to the eigenvalue of the transfer matrix of the 
XXZ model introduced in remark \ref{rm:xxz} inside the restricted 
string hypothesis. Comparing with the known results of 
the XXZ model, it is more precisely the transfer matrix with the 
auxiliary space in the spin $\half$
representation.

The factor in $S_{j,d}$ is proportional to
\begin{equation}
\label{eq:Sc2} 
\prod_{q=1}^{\cD_{j}}
\cK_{1}^{(\hbar_j)}
(\wt \lambda_{j,q}- \wt \lambda_{j,d})
\prod_{q=1}^{\cD_{{j+1}}}
\cK^{(\hbar_{j})}_{\frac{\pi}{\hbar_{j}}-1}
(\wt \lambda_{j+1,q}-\wt \lambda_{j,d})
\prod_{ m\in\cR_{{j}} }
\prod_{\ell=1}^{\wt \nu_{m}}
G^{(\hbar_{j})}_{2(m-\bar s_j)}(\wt 
\lambda_{j,d}-\lambda_{m,\ell})\quad\ 
\end{equation}
Again, this value is similar to the eigenvalue of the transfer matrix 
of the XXZ model introduced in remark \ref{rm:xxz} but now with 
the auxiliary space in the spin $\frac{2\bar s_{j+1}-2\bar s_j-1}{2}$ 
representation.

In addition, the constraint on the type of strings present in 
the model suggests that the underlying model is 
not strictly a XXZ spin chain but rather an RSOS model \cite{bax,ABF} 
(see 
\cite{bare} for the analysis) which is in agreement 
with the counting of states (see section \ref{sec:cou}). Indeed, this 
general case is very similar
to a homogeneous spin $(\bar s_{j+1}-\bar s_{j})$ chain, as treated in 
\cite{NRe} where the underlying RSOS structure has been discovered. 
The only difference lies on the first factor of the L.H.S. of Bethe 
equations (\ref{eq:cBE2})
which may be explained by replacing a homogeneous RSOS model by a 
RSOS model with $\cD_j$ sites with a jump $2\bar s_{j+1}-2\bar 
s_{j}-1$ then $\cD_{j+1}$ sites with a jump $1$.
This interpretation justifies the choice done at 
the end of section \ref{sec:cou} between the different RSOS model 
that are allowed when one looks only at the number of states.

\subsection{Conjecture for the scattering matrix \label{sec:Sm}}

All the considerations of previous subsections \ref{sec:Sg} and \ref{sec:link}
allow us to propose 
an educated guess for the scattering matrix of the model.
This S-matrix depends on the type of particle we consider: the 
particles of type $\cL$ must be treated separately from the particles 
of type $j$ ($1\leq j < \cL$).

\subsubsection{Scattering of type $\cL$ particles}

The particles of type $\cL$ scatter non trivially only with particles 
of the same type and with particles of type $\cL-1$ (if this type 
exists). As explained before, this type of particle has 
a spin $\half$ as well as a supplementary degree of freedom 
satisfying a RSOS model. The non trivial part of the S-matrix for the 
particle of type $\cL$ acts only on
\begin{equation}
 \cH^{RSOS}(\cD_{\cL-1};\cD_{\cL};\bar s_\cL-\bar s_{\cL-1})
\otimes(\CC^2)^{\otimes \cD_\cL}
\end{equation}

We remind (see
 section \ref{sec:link}) that the equations 
(\ref{eq:constL}) and (\ref{eq:Smm}) 
allowing one to compute the part of the scattering matrix acting on 
$(\CC^2)^{\otimes \cD_\cL}$
do not depend on the values of $\bar s_1,\dots,\bar s_\cL$.
Therefore, the spin part of the scattering matrix for the $d^{\text{th}}$
particle of type $\cL$  
is similar to the one of the usual homogeneous spin $\half$ chain 
\cite{faddeev}. As usual, in order to write the scattering matrix, 
we introduce the following transfer matrix
\begin{eqnarray}
t^{{(\cL)}}(\lambda)&=&tr_0 S_{01}(\lambda-\wt 
\lambda_{\cL,1})\dots S_{0\cL}(\lambda-\wt \lambda_{\cL,\cD_\cL})
\label{eq:transfer-spin}
\end{eqnarray}
where 
\begin{equation}
\label{eq:Smat}
S(\lambda)= \cK_1^{(0)}(\lambda)\,R(-\lambda)\,,
\end{equation}
and we have used the R-matrix $R(\lambda)$ defined by (\ref{eq:RR}). 

For the RSOS part of the scattering matrix, we 
use the results of \cite{NRe}. This RSOS part acts on 
$\cH^{RSOS}(\cD_{\cL-1};\cD_{\cL};\bar s_\cL-\bar s_{\cL-1})$.
Let us define the Boltzmann weights of the usual RSOS model 
\cite{bax,ABF} 
as well as the ones of the fused RSOS model 
\cite{djmo,bare} (see also \cite{harpad})
\begin{eqnarray}
W_{\hbar}\left(
\begin{array}{c|c}
d & c\\
\hline
a & b
\end{array}\Big|\lambda
 \right)
&=&
\delta_{ac}
-(-1)^{\frac{a-c}{2}}\,\frac{\sinh(\hbar\lambda)}{\sinh(\hbar(\lambda-i))} 
\sqrt{\frac{\sin(\hbar(a+1))\sin(\hbar(c+1)}{\sin(\hbar(b+1))\sin(\hbar(d+1))}}
\ \delta_{bd}\,,\qquad
\\ 
\label{eq:Wf2}
W_{\hbar}\left(
\begin{array}{c||c}
b_1 & b_{2s}\\
\hline
a_1 & a_{2s}
\end{array}\Big|\lambda
 \right)&=&\sum_{a_2,\dots,a_{2s-1}}\prod_{n=1}^{2s-1}
W_{\hbar}\left(
\begin{array}{c| c}
b_n & b_{n+1}\\
\hline
a_n & a_{n+1}
\end{array}\Big|\lambda+i(n-2s+1)
 \right)\,,
\end{eqnarray}
with $\hbar=\frac{\pi}{2s+2}$ and $2s+2$ the restriction parameter of the RSOS model considered.
The simple lines between the indices in the notation for the Boltzmann weights means that there 
is a jump of $1$ between these ones whereas the double lines means that the jump is $2s-1$.
Let us remark that R.H.S. of (\ref{eq:Wf2}) defining the fused Boltzmann weights are well-defined only 
for $s>\half$. For $s=\half$, the fused Boltzmann weight are $1$.

We introduced the RSOS transfer matrix by its entries, for $0\leq j <\cL$ and 
$1\leq d\leq\cD_{j+1}$,
\begin{eqnarray}
 &&\hspace{-1.5cm}<E(\{a'\}_{\cD_{j}};\{b'\}_{\cD_{j+1}})|
t^{(j,d)}_{1}(\lambda)|
E(\{a\}_{\cD_{j}};\{b\}_{\cD_{j+1}})>=\cN^{(j)}(\lambda)\prod_{q=1}^{\cD_{j}}
W_{\hbar'_{j}}\left(
\begin{array}{c || c}
 a_{q-1} & a_{q}\\
\hline
a_{q}' & a_{q+1}'
\end{array}\Big|\lambda-\wt\lambda_{{j},q}
 \right)
\label{eq:transferRSOS}\\
&&\hspace{2cm}\times
\prod_{q=1}^{d-1}
W_{\hbar'_{j}}\left(
\begin{array}{c|c}
b_{q-1} & b_{q}\\
\hline
b_{q}' & b_{q+1}'
\end{array}\Big|\lambda-\wt\lambda_{{j+1},q}
 \right)
\prod_{q=d}^{\cD_{j+1}-1}
W_{\hbar'_{j}}\left(
\begin{array}{c|c}
b_{q-1} & b_{q}\\
\hline
b_{q}' & b_{q+1}'
\end{array}\Big|\lambda-\wt\lambda_{{j+1},q+1}
 \right)\nonumber
\end{eqnarray}
where $E$ has been defined by (\ref{eq:basrsos}), $\hbar'_{j}=\frac{\pi}{2\bar s_{j+1}-2\bar s_{j}+2}$, $a'_{\cD_j+1}=b_1'$ (by convention) 
and the normalization
\begin{equation}
\cN^{(j)}(\lambda)=
\prod_{q=1}^{\cD_{j}} 
\cK_{2\bar s_{j+1}-2\bar s_{j}-1}^{(\hbar_{j}')}(\wt\lambda_{{j},q}-\lambda)
\prod_{q=1}^{\cD_{j+1}} 
\cK_1^{(\hbar'_{j})}(\wt\lambda_{{j+1},q}-\lambda)\,.
\end{equation}

All these considerations lead to the following conjecture:
\begin{conjecture}\label{con0}
The scattering matrix of the $d^{\text{th}}$ particle of type 
$\cL$ takes the form
\begin{equation}
\label{eq:con_L0}
S_{\cL,d}\sim t^{{(\cL)}}(\wt 
\lambda_{\cL,d})\,t^{{(\cL-1,d)}}_1(\wt \lambda_{\cL,d})
\end{equation}
where $\sim$ stands for 'equals up to a conjugation'.
In (\ref{eq:con_L0}), $t^{{(\cL)}}(\wt \lambda_{\cL,d})$,
acting on $(\CC^2)^{\otimes \cD_\cL}$, is the transfer 
matrix (\ref{eq:transfer-spin})
taken at the value $\lambda=\wt \lambda_{\cL,d}$
and $t^{(\cL-1,d)}_1(\wt \lambda_{\cL,d})$, acting on 
$\cH^{RSOS}(\cD_{\cL-1};\cD_{\cL};\bar s_\cL-\bar s_{\cL-1})$, the 
RSOS transfer matrix (\ref{eq:transferRSOS}) taken at the value 
$\lambda=\wt \lambda_{\cL,d}$.
\end{conjecture}

This conjecture allows us to reproduce the scattering matrices
obtained by \cite{NRe,fafa} 
for the homogeneous spin $s$ chain by putting $\cL=1$ (then $\cD_{\cL-1}=\cD_0=0$),
$\bar s_\cL-\bar 
s_{\cL-1}=s$ and $\hbar'_{\cL-1}=\frac{\pi}{2s+2}$. Obviously, we recover also the 
result for the spin $\half$ chain obtained previously in \cite{faddeev} which is 
the previous case for $s=\half$. In this case, the transfer matrix 
$t^{(0,d)}_1(\lambda)$ must be equal to 1. To show that, we used the following relations
\begin{equation}
\label{eq:props}
\cK_1^{(\pi/3)}(-x)
W_{\pi/3}\left(
\begin{array}{c|c}
0 & 1\\
\hline
1 & 0
\end{array}\Big|x
 \right)=1
\quad\text{and}\quad
\cK_1^{(\pi/3)}(-x)W_{\pi/3}\left(
\begin{array}{c|c}
1 & 0\\
\hline
0 & 1
\end{array}\Big|x
 \right)=1\;.
\end{equation}

In the case where $\bar s_\cL-\bar s_{\cL-1}=1/2$, the RSOS becomes trivial (its dimension is one) and the transfer matrix is reduced to a scalar function
\begin{equation}
\label{eq:SL1L}
 t^{(\cL-1,d)}_1(\wt\lambda_{\cL,d})=
\prod_{q=1}^{\cD_{{\cL-1}}} 
i\coth\left(\frac{\pi}{2}\Big(\wt\lambda_{{\cL},d}-\wt\lambda_{{\cL-1},q}+\frac{i}{2}\Big)\right)\;.
\end{equation}
We have used relations (\ref{eq:props}) and (\ref{eq:exK}).
In particular, we recognize the results obtained in \cite{dev,mene} where the alternating spin $(\half,1)$ chain is treated.

\subsubsection{Scattering for particles of type $1,2,\dots,\cL-1$}

The particles of type $j$ ($1\leq j<\cL$) scatter non trivially only 
with particles of type $j-1$, $j$ and $j+1$. 
The non trivial part of the S-matrix for the particle of type $j$ 
acts only on
\begin{equation}
 \cH^{RSOS}(\cD_{j-1};\cD_{j};\bar s_j-\bar s_{j-1})
\otimes \\
\cH^{RSOS}(\cD_{j};\cD_{j+1};\bar s_{j+1}-\bar s_{j})\;.
\end{equation}
The scattering matrix acting on the first space introduced above is very similar to
the one computed in the previous section and is related to the transfer matrix (\ref{eq:transferRSOS}).

For the scattering matrix acting on the second space, as suggested by relation 
(\ref{eq:Sc2}), we must introduce a transfer matrix with a fused auxiliary space
using the following Boltzmann weights $W_{\hbar}\left(
\begin{array}{c|c}
d & c\\
\hline\hline
a & b
\end{array}\Big|\lambda
 \right)$ and 
$W_{\hbar}\left(
\begin{array}{c||c}
d & c\\
\hline\hline
a & b
\end{array}\Big|\lambda
 \right)$. However, in \cite{bare}, it is shown there exists a gauge transformation
between the Boltzmann weights with a fused auxiliary space and the usual ones.
Therefore,
we introduce the following transfer matrix, for $1\leq j<\cL$ and $1\leq d\leq\cD_j$,
\begin{eqnarray}
&&\hspace{-5mm}<E(\{a'\}_{\cD_{j}};\{b'\}_{\cD_{j+1}})|
t^{(j,d)}_{2s-1}(\lambda)
|E(\{a\}_{\cD_{j}};\{b\}_{\cD_{j+1}})>
=\cM^{(j)}(\lambda) \prod_{q=1}^{\cD_{j+1}}
W_{\hbar'_{j}}\left(
\begin{array}{c|| c}
b_{q-2} & \overline  b_{q-1}\\
\hline
\overline b_{q-1}' & b_{q}'
\end{array}\Big|\lambda-\wt\lambda_{{j+1},q}
 \right)\nonumber\\
&&\hspace{3cm}\times
\prod_{q=1}^{d-1}
W_{\hbar'_{j}}\left(
\begin{array}{c |c}
a_{q-1} & \overline a_{q}\\
\hline
\overline a_{q}' & a_{q+1}'
\end{array}\Big|\lambda-\wt\lambda_{{j},q}
 \right)
\prod_{q=d}^{\cD_{j}-1}
W_{\hbar'_{j}}\left(
\begin{array}{c |c}
a_{q-1} & \overline a_{q}\\
\hline
\overline a_{q}' & a_{q+1}'
\end{array}\Big|\lambda-\wt\lambda_{{j},q+1}
 \right)
\end{eqnarray}
where $\hbar'_{j}=\frac{\pi}{2\bar s_{j+1}-2\bar s_{j}+2}$, $b_{-1}=a_{D_j-1}$, the overlined indices $\overline a=2\bar s_{j+1}-2\bar s_{j}-a$ and the 
normalization
\begin{equation}
\cM^{(j)}(\lambda)=
\prod_{q=1}^{\cD_{j}} \cK_1^{(\hbar_{j}')}(\wt\lambda_{{j},q}-\lambda)
\prod_{q=1}^{\cD_{j+1}} 
\cK_{2\bar s_{j+1}-2\bar s_{j}-1}^{(\hbar'_{j})}(\wt\lambda_{{j+1},q}-\lambda)
\,.
\end{equation}
After these definitions, we can give the conjectured form of 
 the scattering matrix:
\begin{conjecture}
The scattering matrix of the 
$d^{\text{th}}$ particle of type $j$ can be written as follows
\begin{equation} 
\label{con_j}
S_{j,d}\sim t_1^{{(j-1,d)}}(\wt \lambda_{j,d})\,t^{{(j,d)}}_{2(\bar 
s_{j+1}-\bar s_j)-1}(\wt \lambda_{j,d})\;.
\end{equation}
\end{conjecture}

With this conjecture, the scattering matrix $S_{1,d}$ of the alternating spin $(\half,1)$ chain 
reduces to
\begin{equation}
 S_{1,d}\sim \prod_{q=1}^{\cD_{2}} 
i\coth\left(\frac{\pi}{2}\Big(\wt\lambda_{{1},d}-\wt\lambda_{2,q}+\frac{i}{2}\Big)\right)\;.
\end{equation}
This result is consistent with the previous result (\ref{eq:SL1L}) (the scattering of the 
particle 1 on particle 2 must be the same than the scattering of 2 on 1). 
It reproduces also the results of \cite{dev,mene} and, in particular, it shows that 
the particles of type 1 scatter trivially.

Finally, to support this conjecture and the choice of RSOS 
models,
we can look for the central charge of the underlying conformal model 
computed previously in \cite{AlaMart}
\begin{equation}
\label{eq:cc}
c=\cL+\sum_{j=1}^{\cL}\Big(2-\frac{3}{\bar s_i-\bar 
s_{i-1}+1}\Big)\;.
\end{equation}
We recognize in each term $(2-\frac{3}{\bar s_i-\bar s_{i-1}+1})$ the 
central charge
of a RSOS model with the restriction parameter $2\bar s_i-2\bar 
s_{i-1}+2$
(see \cite{bare}).

\subsubsection{Scattering for 2 particles}

Let us remark that for $\cD_j=2$ and $\cD_k=0$ ($k\neq j$), it is 
possible to solve the 
Bethe equations of section \ref{sec:link} and to compute exactly a 
2-particle scattering matrix thanks to the results of 
section \ref{sec:Sg}. However, the results computed in this way disagree, in general, 
with the results obtained via the conjectures (\ref{eq:con_L0}) or (\ref{con_j}).
This discrepancy appeared already in the case of homogeneous spin $s$ chain with $s>\half$:
the 2-body scattering matrices computed in \cite{Tak} are different from the one
computed in \cite{NRe}
(whereas both results agreed for $s=\half$).
This inconsistency has been attributed in \cite{dev,mene} to the 
non-validity of the string hypothesis
(see also the remark \ref{rmk:string}).
To support this point, we emphasize that the computation of 
the central charge has the same feature. Indeed, this computation using the Bethe 
equations inside the string 
hypothesis provides $c=1$ and is different from the one obtained 
by thermodynamical considerations \cite{aff} or by numerical investigations of 
the Bethe equations without string hypothesis \cite{am} which give 
$c=\frac{3s}{s+1}$ (for $s=\half$, both results are again in agreement).
Finally, let us remark that this disagreement occurs when the RSOS structure becomes 
non trivial (i.e. when the RSOS space becomes strictly greater than one).

For the more general case of $L_0$-regular spin chains treated in this paper, we 
have assumed, in order to guess the conjectures, that similar discrepancies in the computation
of the scattering matrix acting non trivially 
on $\cH^{RSOS}(\cD_j;\cD_{j+1};\bar s_{j+1}-\bar s_{j})$ take place
also in the cases when $\bar s_{j+1}-\bar s_{j}>\half$ (i.e. when the corresponding RSOS space 
becomes non trivial) and only in these cases. 
As for the homogeneous spin chain, this is corroborated 
by the computation of the central charge inside the string hypothesis and by comparing it
with (\ref{eq:cc}). Indeed, generalizing the computations done for example in \cite{avdo,vewo,ham}, 
we proved that the central charge is equal to $\cL$ inside the string hypothesis. Therefore,
 to get the value of the central charge given in (\ref{eq:cc}), one must add
a non-vanishing term to $\cL$ each time $s_{j+1}-\bar s_{j}>\half$.

{From} the above discussions, it seems as though the computations done 
in subsections \ref{sec:Sg} and 
\ref{sec:link} are useless. Nevertheless, let us point out the two following points.
Firstly, thanks to the comparison with the homogeneous spin chain results \cite{NRe}, 
they allow us to give an educated guess for the scattering matrices.
Secondly, as argued in \cite{dev}, the scattering matrices obtained inside the string hypothesis 
(i.e. the ones of subsections \ref{sec:Sg} and \ref{sec:link}) may be also the ones of an underlying
quantum field theory. However, this theory would be obtained as a limit 
when one sends to zero first the 
temperature then the magnetic field, whereas the theory corresponding to 
our conjectured scattering
matrices would be obtained when one sends to zero first the magnetic field
and then the temperature.

\section{Conclusion: open problems\label{sec:conclu}}

For a general $L_{0}$-regular closed XXX spin chain, we have 
identified the elementary excitations of the chain. They consist in 
spin-1/2 spinons (associated to the highest spin sites entering the 
chain) and scalar particles, whose different types are related to the other 
different sites in the chain. Then, we have 
conjectured the general form of the scattering matrix for these 
elementary excitations. It makes appear generalized RSOS models, and 
we have given the corresponding Boltzmann weights. 
The first question to address is obviously about the validity of 
this conjecture. We have argued about it by 
computing the central charge of the associated conformal models, but 
a complete proof is still lacking.

It is also natural to ask whether this approach can be generalized to 
other algebras or superalgebras. Indeed, a first account on 
$L_{0}$-regular spin chains based on $gl(N)$ can be found in 
\cite{thermy,therm2}. The relevance of such general integrable spin chains 
in condensed matter physics have been pointed 
out in e.g. \cite{bien}.
However, the calculation of the scattering matrix 
has been done only for homogeneous spin chain with particular 
representations \cite{HJ,DNS}. 
A general treatment remains to be done for these models. This open problem is very 
promising, since it could be linked to RSOS models 
based on $gl(N)$.

The situation is very similar when one considers deformations (quantum 
groups) and/or superalgebras $gl(M|N)$. For the quantum groups $\cU_{q}(gl_{N})$ 
(and in particular $\cU_{q}(gl_{2})$, related to XXZ chain), the same 
algebraic approach to construct integrable $L_0$-regular spin chain can be done 
(see first accounts in \cite{Uqnous,NBA}). 
The excited states for homogeneous arbitrary spin chain based on $\cU_{q}(gl_{2})$ 
and their scattering matrix have been studied in \cite{Kir}: they 
also show internal degrees of freedom. The general case (based on $\cU_{q}(gl_{N})$)
remains to be done. For superalgebras, again the super-Yangian 
$Y(gl(M|N))$ can be investigated through the same method (see e.g. 
\cite{GS}) and should lead to generalized super-RSOS models. 
Finally, $\cU_{q}(gl(M|N))$ can also be treated 
in the same way, see for instance \cite{NBA} where the nested Bethe 
ansatz is done in a unified way for all these cases ($Y(gl_{N})$, 
$\cU_{q}(gl_{N})$, $Y(gl(M|N))$ and $\cU_{q}(gl(M|N))$) and at the 
algebraic level. The computation of the scattering matrices is still an open problem.

For other (orthogonal, symplectic or exceptional) algebras or the 
orthosymplectic superalgebra, the situation is different. 
Indeed, the whole construction of `algebraic spin chains' (as 
introduced in \cite{thermy}) relies on the so-called evaluation 
morphism between the Yangian $Y(gl_{N})$ and the envelopping 
algebra $\cU(gl_{N})$. This morphism does not exist for these other 
classical algebras, so that one needs to work at the level of 
representations directly. 
In fact, the scattering matrices has been computed in few cases (for example, 
in \cite{osp}, for the homogeneous $osp(1|n)$ spin chain in the fundamental 
representation). This indicates that a general method based on
a different approach should exist, but it 
is not known up to now.

Another question that rises is the case of open spin chains. Indeed, 
in \cite{byebye} and \cite{GS} these chains have also been treated, 
and the nested Bethe ansatz for `algebraic open spin chains' based on 
$Y(gl_{N})$, $\cU_{q}(gl_{N})$, $Y(gl(M|N))$ and $\cU_{q}(gl(M|N))$ can be found 
in \cite{OBA}. Thus, we expect that the procedure can be applied to 
these cases too. Scattering matrices for models with boundaries
 when the spins 
are in the fundamental representation have been computed in 
\cite{DNS,anastasia,osp}.

Finally, let us note that one could use these generalized spin chains 
to define new integrable t-J models with impurities in the spirit of 
\cite{tJ}, using a Jordan-Wigner type transformation.

\section*{Acknowledgments:} We thank L. Frappat for discussions and 
encouragements. S.B. wishes to thank N.Y. Reshetikhin for 
fruitful discussions. Part of this work was done when N.C. was in 
SISSA: he thanks them for financial support.
This work was partially supported by the PEPS-PTI
grant \textit{Applications des Mod\`eles Int\'egrables}.


\end{document}